\documentclass[showpacs,amssymb,preprint,preprintnumbers,nofootinbib,superscriptaddress]{revtex4}

\usepackage{amsmath}
\usepackage{graphicx}
\graphicspath{{Figures/}}
\usepackage{latexsym}
\usepackage{amsfonts}
\usepackage{url,hyperref}
\usepackage{bm}
\usepackage{textcomp}
\usepackage{color}
\usepackage{bbm}
\usepackage{slashed}
\usepackage{caption}
\usepackage{epstopdf}
\usepackage{subcaption}
\usepackage{placeins}

\newcommand{\beq}{\begin{equation}}
\newcommand{\eeq}{\end{equation}}
\def\bea{\begin{eqnarray}}
\def\eea{\end{eqnarray}}

\begin{document}
\title{Quasi-normal modes and absorption probabilities of spin-3/2 fields in $D$-dimensional Reissner-Nordstr\"{o}m black hole spacetimes}
\author{C.-H. Chen}
\email[Email: ]{chunhungchen928@gmail.com}
\affiliation{Institute of Physics, Academia Sinica, Nankang, Taipei 11529, Taiwan}
\author{H.~T.~Cho}
\email[Email: ]{htcho@mail.tku.edu.tw}
\affiliation{Department of Physics, Tamkang University, Tamsui District, New Taipei City, Taiwan 25137}
\author{A.~S.~Cornell}
\email[Email: ]{alan.cornell@wits.ac.za}
\affiliation{National Institute for Theoretical Physics; School of Physics, University of the Witwatersrand, Wits 2050, Johannesburg, South Africa}
\author{G. Harmsen}
\email[Email: ]{gerhard.harmsen5@gmail.com}
\affiliation{National Institute for Theoretical Physics; School of Physics, University of the Witwatersrand, Wits 2050, Johannesburg, South Africa}
\author{X. Ngcobo}
\email[Email: ]{540630@students.wits.ac.za}
\affiliation{National Institute for Theoretical Physics; School of Physics, University of the Witwatersrand, Wits 2050, Johannesburg, South Africa}

\begin{abstract}
In this paper we consider spin-3/2 fields in a $D$-dimensional Reissner-Nordstr\"om black hole spacetime. As these spacetimes are not Ricci-flat, it is necessary to modify the covariant derivative to the supercovariant derivative, by including terms related to the background electromagnetic fields, so as to maintain the gauge symmetry. Using this supercovariant derivative we arrive at the corresponding Rarita-Schwinger equation in a charged black hole background. As in our previous works, we exploit the spherically symmetry of the spacetime and use the eigenspinor-vectors on an $N$-sphere to derive the radial equations for both non-transverse-traceless (non-TT) modes and TT modes. We then determine the quasi-normal mode and absorption probabilities of the associated gauge-invariant variables using the WKB approximation and the asymptotic iteration method. We then concentrate on how these quantities change with the charge of the black hole, especially when they reach the extremal limits.
\end{abstract}
\date{\today}
\maketitle


\section{Introduction}

\par In supergravity theories \cite{dasfre,gripen} the gravitino is described by a spin-3/2 field. The equations of motion of these spin-3/2 fields are given by the Rarita-Schwinger equation:
\begin{equation}
\gamma^{\mu\nu\alpha}\nabla_{\nu}\psi_{\alpha}=0,
\end{equation}
where 
\begin{equation}
\gamma^{\mu\nu\alpha}\equiv\gamma^{[\mu}\gamma^{\nu}\gamma^{\alpha]}=\gamma^{\mu}\gamma^{\nu}\gamma^{\alpha}-\gamma^{\mu}g^{\nu\alpha}+\gamma^{\nu}g^{\mu\alpha}-\gamma^{\alpha}g^{\mu\nu}
\end{equation}
is antisymmetric product of Dirac gamma matrices, $\nabla_{\nu}$ is the covariant derivative, and $\psi_{\alpha}$ is the spin-3/2 field.
In four dimensional black hole spacetimes the Rarita-Schwinger equations are usually analyzed in the Newman-Penrose formalism. However, this formalism cannot be extended to higher dimensions in a straightforward way. In our previous works \cite{Chen2015,Chen2016} we have tried an alternative approach to deal with spherically symmetric black hole cases. Using a complete set of eigenspinor-vectors on $N$-spheres, we were able to separate the radial and angular parts of the Rarita-Schwinger equation. In this paper we would like to extend our considerations to charged black hole spacetimes.

\par The Rarita-Schwinger equation is invariant under the gauge transformation
\begin{equation}
\psi'_{\alpha}=\psi_{\alpha}+\nabla_{\alpha}\varphi,\label{gauge}
\end{equation}
where $\varphi$ is a gauge spinor, provided that the background spacetime is Ricci-flat \cite{Chen2015,Chen2016}. This is not the case for charged black holes, nor for black holes in de Sitter or anti-de Sitter spaces. To maintain the gauge symmetry in those cases it is necessary to modify the covariant derivative into the so-called the ``supercovariant derivative". This is done by adding terms related to the cosmological constant and the electromagnetic field of the black hole. Here we shall concentrate on charged Reissner-Nordstr\"om black holes in asymptotically flat spacetimes, where in the following section we shall show in detail how the supercovariant derivative is constructed in this case. 

\par Using the supercovariant derivative we are able to obtain the Rarita-Schwinger equation for spin-3/2 fields in Reissner-Nordstr\"om black hole spacetimes. Since the spacetime is still spherically symmetric, it is possible, as in our previous works, to derive the radial equations for each component of the spin-3/2 field using eigenspinor-vectors on the $N$-sphere. However, the component fields are not gauge invariant, while the physical fields should be. Hence we shall, as in Ref. \cite{Chen2016}, construct a combination of the component fields, which is gauge invariant. That is, we shall use the same gauge invariant variables and work out the corresponding radial equations. 

\par As the aim in this paper is to study spin-3/2 fields near a Reissner-Nordstr\"{o}m black hole, we will focus on how the charge Q of the black hole affects the behavior of the fields. This is done by studying the quasi-normal models (QNMs) associated to our fields, where QNMs are characterized by their complex frequencies. The real parts of the frequencies represent the frequencies of oscillations, while the imaginary parts the decay constants of damping. These QNMs are uniquely determined by the parameters of the black hole \cite{konzen}, where in order to determine these QNMs we will use the WKB and improved Asymptotic Iterative Method (AIM), where these methods and how to implement them are given in Refs. \cite{kon,chocor1,Chen2016}. Finally, using the WKB method we are able to obtain the absorption probabilities associated to our spin-3/2 fields, which can give us an insight into the grey-body factors and cross-sections of the black hole.

\par As such, this paper is set out as follows: In the next section we give a brief motivation for the form of our supercovariant derivative, and in Sec. \ref{Sec:Pot} we use this supercovariant derivative to determine our equations of motion, and the potential functions for both the ``non-TT'' and the ``TT eigenmodes'' of our fields. In Sec. \ref{Sec:QNMs}, we present the QNMs for our spin-3/2 fields near the Reissner-Nordstr\"{o}m black hole. The corresponding absorption probabilities are laid out in Sec. \ref{Sec:Abs}. Finally, in Sec. \ref{Sec:Conclusion}, we give concluding remarks on our results.


\section{Supercovariant derivative}\label{Sec:CoDev}

\par In order to ensure that our Rarita-Schwinger equation, $\gamma^{\mu\nu\alpha}\nabla_{\nu}\psi_{\alpha}=0$, remains true, we must require that our spinor-vectors, $\psi_{\mu}$, are invariant under the transformation in Eq.~(\ref{gauge}). This is guaranteed if
\begin{equation}\label{gauge symmetry}
\frac{1}{2}\gamma^{\mu\nu\alpha}\left[\nabla_{\nu},\nabla_{\alpha}\right]\varphi=0,
\end{equation}
which is satisfied when the metric is a Ricci flat spacetime \cite{Chen2016}. However, in the case of the Reissner-Nordstr\"{o}m metric this is not necessarily true, and so we must first determine the supercovariant derivative, $\tilde{\mathcal{D}}_\mu$. We make the assumption that the derivative has the form:
\begin{equation}\label{Supercovariantsetting}
\tilde{\mathcal{D}}_{\mu}=\mathcal{D}_{\mu}+b\gamma^{\rho}F_{\mu\rho}+c\gamma_{\mu\rho\sigma}F^{\rho\sigma},
\end{equation}
which needs to satisfy:
\begin{equation}\label{gauge symmetry1}
\frac{1}{2}\gamma^{\lambda\mu\nu}\left[\tilde{\mathcal{D}}_{\mu},\tilde{\mathcal{D}}_{\nu}\right]\varphi=0,
\end{equation}
where $\mathcal{D}_{\mu}=\nabla_{\mu}-ieA_{\mu}$, $F_{\mu\nu}$ is the electromagnetic field, and $b$, $c$ are unknown constants. Plugging Eq. (\ref{Supercovariantsetting}) into Eq. (\ref{gauge symmetry1}) we have:
\begin{equation}\label{scd1}
\begin{aligned}
0= & 4c\left(D-2\right)\left(\nabla_{\mu}F^{\mu\lambda}\right)\varphi \\
   & +\gamma^{\mu}[G_{\mu}^{\ \lambda}-4\left(b^{2}+2c\left(D-3\right)\right)F_{\mu\nu}F^{\nu\lambda}-2\left(b^{2}-2c^{2}\left(D-3\right) \left(D-4\right)\right)g_{\mu}^{\ \lambda}F_{\rho\sigma}F^{\rho\sigma}]\varphi \\
   & +\gamma^{\mu\nu}[ \left(b+2c\left(D-3\right)\right)\left(2g_{\mu}^{\ \lambda}\nabla_{\alpha}F^{\alpha}_{\ \nu}-\nabla^{\lambda}F_{\mu\nu}\right)]\varphi \\
   & +\gamma^{\mu\rho\sigma}[-ieg_{\mu}^{\ \lambda}F_{\rho\sigma}+ 4\left(b+2c\left(D-3\right)\right)\left(b+c\left(D-6\right)\right)F_{\rho\sigma}F_{\mu}^{\ \lambda}]\varphi \\
   &+\gamma^{\lambda\mu\nu\rho\sigma}[-2\left(b^{2}+2bc\left(D-5\right)+c^{2}\left(D^{2}-11D+26\right)\right)]F_{\mu\nu}F_{\rho\sigma}\varphi.
\end{aligned}
\end{equation}
Setting $\gamma^{\mu\nu}$ equal to zero we have $b=-2(D-3)c$, and together with Eq. (\ref{scd1}) we have:
\begin{equation}\label{scd2}
\begin{aligned}
0= & 4c\left(D-2\right)\left(\nabla_{\mu}F^{\mu\lambda}\right)\varphi \\
   & +\gamma^{\mu}\Bigg[G_{\mu}^{\ \lambda}+16\left(D-2\right)\left(D-3\right)c^{2}\left(F_{\mu\nu}F^{\lambda\nu}-\frac{1}{4}g_{\mu}^{\ \lambda}F_{\rho\sigma}F^{\rho\sigma}\right)\Bigg]\varphi \\
   & +\gamma^{\mu\rho\sigma}[-ieg_{\mu}^{\ \lambda}F_{\rho\sigma}] +\gamma^{\lambda\mu\nu\rho\sigma}\left(-2c^{2}\right)\left(D-1\right)\left(D-2\right)F_{\mu\nu}F_{\rho\sigma}\varphi.
\end{aligned}
\end{equation}
In order to remove the $\gamma^{\mu}$ terms we require that
\begin{equation}
\begin{aligned}
G_{\mu}^{\ \lambda}+16\left(D-2\right)\left(D-3\right)c^{2}\left(F_{\mu\nu}F^{\lambda\nu}-\frac{1}{4}g_{\mu}^{\ \lambda}F_{\rho\sigma}F^{\rho\sigma}\right)=0,
\end{aligned}
\end{equation}
where
\begin{equation}\label{kappa}
16\left(D-2\right)\left(D-3\right)c^{2}=-\frac{1}{2} \ \ \Rightarrow \ \ c=\frac{i}{4\sqrt{2\left(D-2\right)\left(D-3\right)}},
\end{equation}
and with the $\gamma^{\mu\rho\sigma}$ term equal to zero only when $e=0$.  Next we consider the $\gamma^{\mu}$ and $\gamma^{\lambda\mu\nu\rho\sigma}$ terms. In the four dimensional case, $\gamma^{\lambda\mu\nu\rho\sigma}=0$ and $\nabla_{\mu}F^{\mu\lambda}=0$, which is the Maxwell equation, and Eq. (\ref{scd2}) is automatically satisfied.  In the five dimensional case, $\gamma^{\lambda\mu\nu\rho\sigma}$ is proportional to the identity matrix, such that we have to set the equations of motion for the electromagnetic field (where $D=5$) as:
\begin{equation}
\nabla_{\mu}F^{\mu\lambda}=-\frac{1}{4\sqrt{3}}\epsilon^{\lambda\mu\nu\rho\sigma}F_{\mu\nu}F_{\rho\sigma} \ \ , \ \ \epsilon^{tr\theta_{1}\theta_{2}\theta_{3}}=\frac{1}{\sqrt{-g}},
\end{equation}
where $\epsilon^{\lambda\mu\nu\rho\sigma}$ is the Levi-Civita tensor. In higher dimensional cases we can take the equations of motion for electromagnetic field to be $\nabla_{\mu}F^{\mu\lambda}=0$, and  the $\gamma^{\lambda\mu\nu\rho\sigma}$ term vanishes if the condition $F_{\mu[\nu}F_{\rho\sigma]}=0$ is fulfilled. Finally the supercovariant derivative for the spin-3/2 field in a general dimensional Reissner-Nordstr\"{o}m black hole spacetime can be written as:
\begin{equation}\label{Supercovariantd}
\tilde{\mathcal{D}}_{\mu}=\nabla_{\mu}+\frac{1}{2}\sqrt{\frac{D-3}{2\left(D-2\right)}}\gamma_{\rho}F_{\mu}^{\ \rho}+\frac{i}{4\sqrt{2\left(D-2\right)\left(D-3\right)}}\gamma_{\mu\rho\sigma}F^{\rho\sigma}.
\end{equation}
Note that this is consistent with the results of Ref. \cite{Liu2014}, though we must emphasize that in our construction one cannot find an appropriate supercovariant derivative for a ``charged" spin-3/2 field in the Reissner-Nordstr\"{o}m black hole spacetime.


\section{Potential function}\label{Sec:Pot}

\par In this section we will determine both the radial equations and the potential functions for our spin-3/2 fields near the Reissner-Nordstr\"{o}m black hole, using the same approach as we have done for the $N$ dimensional Schwarzschild black hole. For completeness we reproduce some of the results from the Schwarzschild case in this paper, where a full explanation of this method can be found in Ref.\cite{Chen2016}.


\subsection{Rarita-Schwinger field near D dimensional Reissner-Nordstr\"{o}m black holes}

\par Our line element is given as \cite{Liu2014}:
\begin{equation}
\begin{aligned}
ds^{2}=-fdt^{2}+\frac{1}{f}dr^{2}+r^{2}d\bar{\Omega}^{2}_{N},
\end{aligned}
\end{equation}
where $f=1-\frac{2M}{r^{D-3}}+\frac{Q^{2}}{r^{(2D-6)}}$  and $D=N+2$. The term $d\bar{\Omega}_{N}$ denotes the metric of the $N$ sphere $S^{N}$, where we will use over-bars to represent terms from this metric. Next, the electromagnetic field takes the Coulomb form which is given as \cite{KodaIshi2004}:
\begin{equation}
\begin{aligned}
F_{tr}=\frac{q}{r^{(D-2)}}\ \ , \ \ A_t=\frac{q}{(D-3)r^{(D-3)}}.
\end{aligned}
\end{equation}
The relation of $Q$ and $q$ is
\begin{equation}
\begin{aligned}
Q^{2}=\frac{\kappa^{2}q^{2}}{(D-2)(D-3)},\label{qQ}
\end{aligned}
\end{equation}
where $\kappa^{2}$ is a constant defined by the Einstein field equation,
\begin{equation}
\begin{aligned}
G_{\mu\nu}=\kappa^{2}\left(F_{\mu\lambda}F_{\nu}^{\ \lambda}-\frac{1}{4}g_{\mu\nu}F_{\rho\sigma}F^{\rho\sigma}\right).
\end{aligned}
\end{equation}
In order to be consistent with the supercovariant derivative above, one should take $\kappa^{2}=1/2$ as in Eq.~(\ref{kappa}). Since we represent the wave functions of our fields as spinor-vectors, which  can be constructed from the ``non TT eigemodes'' and the ``TT eigenmodes'' on $S^{N}$, we will use the massless form of the Rarita-Schwinger equation \cite{Chen2016}:
\begin{equation}\label{Rarita-Schwinger}
\begin{aligned}
\gamma^{\mu\nu\alpha}\tilde{\mathcal{D}}_{\nu}\psi_{\alpha}=0,
\end{aligned}
\end{equation}
where $\tilde{\mathcal{D}}_{\nu}$ is the supercovariant derivative in Eq.~(\ref{Supercovariantd}). 


\subsection{Non-TT eigenfunctions}

\par The radial and temporal wave functions can be written as:
\begin{equation}\label{DefphiRphit}
\begin{aligned}
\psi_{r}=\phi_{r}\otimes\bar{\psi}_{(\lambda)} \:\:\: \text{and} \:\:\: \psi_{t}=\phi_{t}\otimes\bar{\psi}_{(\lambda)},
\end{aligned}
\end{equation}
where $\bar{\psi}_{(\lambda)}$ is an eigenspinor on $S^{N}$, with eigenvalue $i\bar{\lambda}$. The eigenvalues $\bar{\lambda}$ are given by $\bar{\lambda}=\left(j+(D-3)/2 \right)$, where $j=3/2,5/2,7/2,...$ \cite{Chen2016}. Our angular wave function is written as:
\begin{equation}
\begin{aligned}
\psi_{\theta_{i}}=\phi_{\theta}^{(1)}\otimes\bar{\nabla}_{\theta_{i}}\bar{\psi}_{(\lambda)}+\phi^{(2)}_{\theta}\otimes\bar{\gamma}_{\theta_{i}}\bar{\psi}_{(\lambda)},
\end{aligned}
\end{equation}
where $\phi_{\theta}^{(1)}$, $\phi_{\theta}^{(2)}$ are functions of $r$ and $t$ which behave like 2-spinors. We begin by using the Weyl gauge, $\phi_{t}=0$, and then introduce a gauge invariant variable, which we also use to determine the equations of motion. Looking at $\mu=r$, $\mu=t$ and $\mu=\theta$ equations in Eq. (\ref{Rarita-Schwinger}) separately, we determine the 4 appropriate equations of motion. Note that our choice of the gamma tensors and the spin connections can be found in Ref. \cite{Chen2016}.


\subsubsection{Equations of motion}

\par Firstly consider the case of $\mu=t$ in Eq. (\ref{Rarita-Schwinger}):
\begin{equation}
\begin{aligned}
\gamma^{t\nu\alpha}\tilde{\mathcal{D}}_{\nu}\psi_{\alpha}=0.
\end{aligned}
\end{equation}
Using the definitions of our wave functions we determine our first equation of motion to be:
\begin{equation}\label{NonTTEoM1}
\begin{aligned}
0=&-\Bigg(i\bar{\lambda} + (D-2)\frac{\sqrt{f}}{2}i\sigma^{3} +  \left(D-2 \right)\frac{iQ}{2r^{D-3}}\Bigg)\phi_{r}\\
&+ \Bigg(i\bar{\lambda}\partial_{r} -\frac{1}{4}\frac{\left(D-2\right)\left(D-3\right)}{r\sqrt{f}}i\sigma^{3} + (D-3)\frac{i\bar{\lambda}}{2r} \Bigg)\phi^{(1)}_{\theta}\\
&+ \Bigg((D-2)\partial_{r} + (D-3)\frac{i\bar{\lambda}}{r\sqrt{f}}i\sigma^{3} + \frac{(D-2)(D-3)}{2r}\Bigg)\phi^{(2)}_{\theta}.\\
\end{aligned}
\end{equation}
Next we consider $\mu=r$, and get the second equation of motion as:
\begin{equation} \label{NonTTEoM2}
\begin{aligned}
0 = & \Bigg[-\frac{i\bar{\lambda}}{\sqrt{f}}\partial_{t}+\frac{i\bar{\lambda} f'}{4\sqrt{f}}\sigma^{1}-\frac{(D-3)(D-2)}{4r}\sigma^{2} + (D-3)\frac{i\bar{\lambda}\sqrt{f}}{2r}\sigma^{1}\Bigg]\phi^{(1)}_{\theta} \\
&+ \Bigg[-\frac{D-2}{\sqrt{f}}\partial_{t}+\frac{(D-2)f'}{4\sqrt{f}}\sigma^{1}+(D-3)\frac{i\bar{\lambda}}{r}\sigma^{2} + (D-2)(D-3)\frac{\sqrt{f}}{2r}\sigma^{1} \Bigg]\phi^{(2)}_{\theta}.\\
\end{aligned}
\end{equation}
Finally for the case of $\mu=\theta_{i}$ we obtain two more equations:
\begin{equation}
\begin{aligned}
0 = & \left(\partial_{t} - \frac{f'}{4}\sigma^{1}+ i\bar{\lambda}\frac{\sqrt{f}}{r}\sigma^{2}- (D-3)\frac{f}{2r}\sigma^{1}\right)\phi_{r}\\
& + \Bigg(\frac{\bar{\lambda}}{r\sqrt{f}}\sigma^{3}\partial_{t} - i\bar{\lambda}\frac{f'}{4r\sqrt{f}}\sigma^{2} -i\bar{\lambda}\frac{\sqrt{f}}{r}\sigma^{2}\partial_{r} -\frac{\left(D-3\right)\left(D-4\right)}{4r^{2}}\sigma^{1} - i\bar{\lambda}(D-4)\frac{\sqrt{f}}{2r^{2}}\sigma^{2}\\
&\hspace{1cm} -\bar{\lambda}\left(D-2\right)\frac{Q}{2r^{D-1}}\sigma^{1}\Bigg)\phi^{(1)}_{\theta} \\
& + \Bigg(- \frac{D-3}{r\sqrt{f}}i\sigma^{3}\partial_{t} - (D-3)\frac{f'}{4r\sqrt{f}}\sigma^{2} - (D-3)\frac{\sqrt{f}}{r}\sigma^{2}\partial_{r} + (D-4)\frac{i\bar{\lambda}}{r^{2}}\sigma^{1}\\
&  \hspace{1cm} - (D-3)(D-4)\frac{\sqrt{f}}{2r^{2}}\sigma^{2} +\left(D-3 \right)\left(D-2\right)\frac{iQ}{2r^{D-1}}\sigma^{1}\Bigg)\phi^{(2)}_{\theta} \\
\end{aligned}
\end{equation}
\begin{equation}\label{NonTTEoM3}
\begin{aligned}
0 & = -\frac{\sqrt{f}}{r}\sigma^{2}\phi_{r}  + \Bigg(\frac{1}{r\sqrt{f}}i\sigma^{3}\partial_{t} + \frac{f'}{4r\sqrt{f}}\sigma^{2} + \frac{\sqrt{f}}{r}\sigma^{2}\partial_{r} + (D-4)\frac{\sqrt{f}}{2r^{2}}\sigma^{2}\\
&\hspace{2cm} - \left(D-2 \right)\frac{iQ}{2r^{D-1}}\sigma^{1}\Bigg)\phi^{(1)}_{\theta} - \frac{D-4}{r^{2}}\sigma^{1}\phi_{\theta}^{(2)}.
\end{aligned}
\end{equation}
It can be shown that these four equations of motion are not independent. One of them can be obtained from the other three. Hence we shall work with only Eqs. (\ref{NonTTEoM1}), (\ref{NonTTEoM2}) and (\ref{NonTTEoM3}) in the following.


\subsubsection{Effective potential}

\par The functions $\phi_{r}$, $\phi_{\theta}^{(1)}$ and $\phi_{\theta}^{(2)}$ are not gauge invariant, and as such we apply a gauge invariant variable to our equations of motion. Using the same arguments as we have used in Ref.\cite{Chen2016} we obtain the following gauge invariant variable
\begin{equation}
\begin{aligned}
\Phi = -\left(\frac{\sqrt{f}}{2}i\sigma^{3} +\frac{iQ}{2r^{D-3}} \right)\phi^{(1)}_{\theta} +\phi^{(2)}_{\theta}.
\end{aligned}
\end{equation}
Plugging this into Eq. (\ref{NonTTEoM1}), Eq. (\ref{NonTTEoM2}) and Eq. (\ref{NonTTEoM3}) we obtain the equation of motion for the gauge invariant variable $\Phi$,
\begin{equation}\label{nonTTGaugeIn}
\begin{aligned}
& \left((D-2)\frac{\sqrt{f}}{2} +\left(\bar{\lambda}+ \left(D-2 \right)\frac{Q}{2r^{D-3}}\right)\sigma^{3}\right)\\
&\times \Bigg[-\frac{D-2}{f}\sigma^{1}\partial_{t}+(D-2)\frac{f'}{4f}-\frac{(D-3)\bar{\lambda}}{r\sqrt{f}}\sigma^{3} + \frac{(D-2)(D-3)}{2r} \Bigg]\Phi =\\
& \hspace{6cm} \left((D-2)\frac{\sqrt{f}}{2}-\left(\bar{\lambda} + \left(D-2 \right)\frac{Q}{2r^{D-3}}\right)\sigma^{3}\right)\\
&\times \Bigg[(D-2)\partial_{r} - \frac{\bar{\lambda}}{r\sqrt{f}}\sigma^{3} + \frac{(2D-7)(D-2)}{2r} + \left(D-2 \right)(D-4)\frac{Q}{2r^{D-2}\sqrt{f}}\sigma^{3}\Bigg]\Phi.
\end{aligned}
\end{equation}
Component-wise, $\Phi$ can be written as:
\begin{equation}\label{DefPhi}
\begin{aligned}
\Phi =
\begin{pmatrix}
\phi_{1}e^{-i\omega t}\\
\phi_{2}e^{-i\omega t}\\
\end{pmatrix},
\end{aligned}
\end{equation}
where $\phi_{1}$ and $\phi_{2}$ are purely radially dependent terms. Furthermore we set
\begin{equation}\label{Defphi}
\begin{aligned}
\phi_{1}=\frac{(\frac{D-2}{2})^{2}f-\left(\bar{\lambda}+C \right)^{2}}{Br^{\frac{D-4}{2}}f^{1/4}}\tilde{\phi}_{1}\:\:\: \text{and} \:\:\: \phi_{2}=\frac{(\frac{D-2}{2})^{2}f-\left(\bar{\lambda}-C \right)^{2}}{Ar^{\frac{D-4}{2}}f^{1/4}}\tilde{\phi}_{2},
\end{aligned}
\end{equation}
where
\begin{equation}
\begin{aligned}
A &= \frac{D-2}{2}\sqrt{f} + \left( \bar{\lambda} + C\right), \:\:\:B = \frac{D-2}{2}\sqrt{f} -\left( \bar{\lambda} + C \right)\:\:\: \text{and} \:\:\:C = \left(D-2 \right)\frac{Q}{2r^{D-3}}.
\end{aligned}
\end{equation}
Applying Eq. (\ref{DefPhi}) and Eq. (\ref{Defphi}) to Eq. (\ref{nonTTGaugeIn}) we get the following set of coupled equations:
\begin{equation}
\begin{aligned}
\left(f\partial_{r}-W \right)\tilde{\phi}_{1} & =i\omega\tilde{\phi}_{2} , \\
\left(f\partial_{r}+W \right)\tilde{\phi}_{2} & =i\omega\tilde{\phi}_{1},
\end{aligned}
\end{equation}
where
\begin{equation}
\begin{aligned}
W = & \frac{(D-3)\sqrt{f}}{rAB}\Bigg[( \bar{\lambda} + C)\frac{2}{D-2}AB + \frac{D-2}{2}\left(C + \bar{\lambda}\left(1 - f\right)\right) \Bigg]\\
& \hspace{1cm} - \frac{D-4}{r(D-2)}\sqrt{f}(\bar{\lambda}+ C).
\end{aligned}
\end{equation}
Decoupling these two equations we obtain the following radial equations
\begin{equation}\label{nonTTPot}
\begin{aligned}
-\frac{d^{2}}{dr_{*}^{2}}\tilde{\phi}_{1} + V_{1}\tilde{\phi}_{1} & = \omega^{2}\tilde{\phi}_{1}, \\
-\frac{d^{2}}{dr_{*}^{2}}\tilde{\phi}_{2} + V_{2}\tilde{\phi}_{2} & = \omega^{2}\tilde{\phi}_{2},
\end{aligned}
\end{equation}
where $r_{*}$ is the tortoise coordinate with the definition $dr_{*}=\frac{1}{f(r)}dr$, and 
$$V_{1,2} = \pm f(r)\frac{dW}{dr} + W^{2}.$$ 
Setting $Q=0$ in Eqs. (\ref{nonTTPot}) we recover the Schwarzschild potential as given in Ref. \cite{Chen2016}.


\subsection{TT eigenfunctions}


\subsubsection{Equations of motion}

\par Setting the $\psi_{r}$ and $\psi_{t}$ to be the same as in the ``non-TT eigenfunctions'' case given in Eq. (\ref{DefphiRphit}), the angular part is now given as:
\begin{equation}
\begin{aligned}
\psi_{\theta_{i}}=\phi_{\theta}\otimes\bar{\psi}_{\theta_{i}},
\end{aligned}
\end{equation}
where $\bar{\psi}_{\theta_{i}}$ is the TT mode eigenspinor-vector which includes the ``TT mode I'' and ``TT mode II'', as described in Ref. \cite{Chen2016}, and $\phi_{\theta}$ behaves like a 2-spinor. We again initially use the Weyl gauge, and in this case apply the TT conditions on a sphere, giving us $\phi_{t}=\phi_{r}=0$ \cite{Chen2016}. Our only non-zero equation of motion is then determined to be
\begin{equation}\label{EoMTT}
\begin{aligned}
&\Bigg(\frac{1}{r\sqrt{f}}i\sigma^{3}\partial_{t} + \frac{\sqrt{f}}{r}\sigma^{2}\partial_{r} + \frac{f'}{4r\sqrt{f}}\sigma^{2} + \left(D-4 \right)\frac{\sqrt{f}}{2r^{2}}\sigma^{2} + \frac{i\bar{\zeta}}{r^{2}}\sigma^{1} -\left(D-2 \right)\frac{iQ}{2r^{D-1}}\sigma^{1} \Bigg)\phi_{\theta} = 0,
\end{aligned}
\end{equation}
where in this case $\phi_{\theta}$ is already gauge invariant. We can therefore use this equation to determine our radial equation.


\subsubsection{Effective potential}

\par We can rewrite $\phi_{\theta}$ as
\begin{equation}\label{PsiDefTT}
\begin{aligned}
\phi_{\theta}=\sigma^{2}
\begin{pmatrix}
\Psi_{\theta_{1}}e^{-i\omega t}\\
\Psi_{\theta_{2}}e^{-i\omega t}\\
\end{pmatrix},
\end{aligned}
\end{equation}
and substituting Eq.(\ref{PsiDefTT}) into Eq. (\ref{EoMTT}) we get the following set of coupled equations:
\begin{equation}\label{CoupledTTEoM}
\begin{aligned}
& \left(f\partial_{r} + \frac{f'}{4}+\left(D-4 \right)\frac{f}{2r} -\left(\frac{\bar{\zeta}\sqrt{f}}{r}-\left(D-2 \right)\frac{Q\sqrt{f}}{2r^{D-2}} \right)\right)\Psi_{\theta_{1}} = i\omega\Psi_{\theta_{2}},\\
& \left(f\partial_{r} + \frac{f'}{4}+\left(D-4 \right)\frac{f}{2r} +\left(\frac{\bar{\zeta}\sqrt{f}}{r}-\left(D-2 \right)\frac{Q\sqrt{f}}{2r^{D-2}} \right)\right)\Psi_{\theta_{2}} = i\omega\Psi_{\theta_{1}}.
\end{aligned}
\end{equation}
Setting
\begin{equation}
\begin{aligned}
\tilde{\Psi}_{\theta_{1}}=r^{\frac{D-4}{2}}f^{\frac{1}{4}}\Psi_{\theta_{1}}\,\,\, \text{and} \,\,\, \tilde{\Psi}_{\theta_{2}}=r^{\frac{D-4}{2}}f^{\frac{1}{4}}\Psi_{\theta_{2}},
\end{aligned}
\end{equation}
we can simplify the equations in Eq. (\ref{CoupledTTEoM}), and get the following
\begin{equation}\label{TTCoupled}
\begin{aligned}
&\left(f\partial_{r}-\mathbb{W} \right)\tilde{\Psi}_{\theta_{1}}=i\omega\tilde{\Psi}_{\theta_{1}},\\
&\left(f\partial_{r} + \mathbb{W} \right)\tilde{\Psi}_{\theta_{2}}=i\omega\tilde{\Psi}_{\theta_{2}},\\
\end{aligned}
\end{equation}
where
\begin{equation}
\begin{aligned}
\mathbb{W}=\frac{\bar{\zeta}\sqrt{f}}{r}-\left(D-2 \right)\frac{Q\sqrt{f}}{2r^{D-2}}.
\end{aligned}
\end{equation}
We now decouple the equations in Eq. (\ref{TTCoupled}) and obtain the radial equations
\begin{equation}
\begin{aligned}
-\frac{d}{dr_{*}^{2}}\tilde{\Psi}_{\theta_{1}}+\mathbb{V}_{1}& =\omega^{2}\Psi_{\theta_{1}} \\ 
\text{and}\:\:\: -\frac{d^{2}}{dr_{*}^{2}}\tilde{\Psi}_{\theta_{2}} + \mathbb{V}_{2}\tilde{\Psi}_{\theta_{2}} & = \omega^{2}\tilde{\Psi}_{\theta_{2}},
\end{aligned}
\end{equation}
where 
$$\mathbb{V}_{1,2} = \pm f(r)\frac{d\mathbb{W}}{dr} + \mathbb{W}^{2}$$ 
and our eigenvalue $\bar{\zeta}$ is given as $\bar{\zeta}=j+(D-3)/2$ with $j=3/2,5/2,7/2,...$. 

\par As noted in Ref. \cite{Chen2016}, the Schwarzschild case of this potential is the same as for Dirac particles in a general dimensional Schwarzschild black hole \cite{cho2007split}. This, however, is not true for the Reissner-Nordstr\"{o}m case. For the spin-3/2 field one needs to use the supercovariant derivative in the charged black hole spacetime, whereas for the Dirac field one would still use the ordinary covariant derivative. The extra terms in the supercovariant derivative would render the effective potential of the spin-3/2 field in the TT mode to be different from that of the Dirac field in the same spacetime.


\section{QNMs}\label{Sec:QNMs}

\par In order to obtain the QNMs we have chosen to use the WKB method, to 3rd and 6th order, and AIM. We would like to investigate how the quasi-normal frequencies change with the charge $Q$ of the black hole, where a particularly interesting case of the Reissner-Nordstr\"om black hole would be the extremal case $Q=M$. In this section we present the QNMs in the cases $Q=0.1M$, $Q=0.5M$, and $Q=M$ for both  ``non-TT eigenfunction related" and ``TT eigenfunction related" potentials from $D=4$ to $D=7$.


\subsection{non-TT eigenfunctions related}

\par In this subsection we consider the QNMs of the radial equations in Eq.~(\ref{nonTTPot}). Since $V_{1}$ and $V_{2}$ are isospectral, we can choose either one to work with. Here we shall concentrate on the first equation, that is, the potential $V_{1}$. In the case of the WKB methods for the QNMs associated to the ``non-TT eigenfunctions'', the full explanation of how to determine the QNMs Refs. \cite{iyewil,kon}. For the AIM, a detailed discussion can be found in Refs. \cite{chocor1,Chen2015, Chen2016}. 

\par For the case of the Reissner-Nordstr\"{o}m background, we shall start with the definition of the tortoise coordinate
\begin{eqnarray}
& \displaystyle dr_{*}=\frac{1}{f(r)}dr, & \notag \\ 
& \displaystyle f(r)=1-\frac{2M}{r^{(D-3)}}+\frac{Q^{2}}{r^{2(D-3)}}=\frac{(r^{(D-3)}-r_{+})(r^{(D-3)}-r_{-})}{r^{2(D-3)}},&
\end{eqnarray}
where $r_{\pm}=M\pm\sqrt{M^{2}-Q^{2}}$. The relation between $r$ and $r_{*}$ can be obtained by
\begin{eqnarray}
r_{*} & = & \int dr+\int \frac{2r_{+}-Q^{2}}{(r_{+}-r_{-})(r^{D-3}-r_{-})} dr+\int \frac{-2r_{-}+Q^{2}}{(r_{+}-r_{-})(r^{D-3}-r_{+})}dr , \; \mathrm{for} \ \ Q<M , \nonumber\\ \\
\mathrm{and} \ \ r_{*} & = & \int \frac{r^{2(D-3)}}{r^{2(D-3)}-2r^{D-3}+1}dr\ , \; \mathrm{for} \ \ Q=M.
\end{eqnarray}
In the AIM we first single out the asymptotic behavior of $\tilde{\phi}_{1}$, which is due to the QNM boundary conditions, 
\begin{eqnarray}
\tilde{\phi}_{1}\rightarrow e^{\pm i \omega r_{*}}\sim\alpha \ \ , \ \ r\rightarrow \pm \infty.
\end{eqnarray}
Note that we use ``$\sim$" to signify that for convenience we choose the asymptotic function $\alpha$ to include the leading term and some of the subleading terms of $e^{\pm i \omega r_{*}}$, but not necessarily the whole exponential. Next, a necessary coordinate transformation in the AIM will be
\begin{equation}
\xi^{2}(r)=1-\frac{(r_{+})^{\frac{1}{D-3}}}{r}.
\end{equation}
The coefficients for the lowest order can be obtained as:
\begin{eqnarray}
\lambda_{0}& = & -\left(\frac{f'}{f}+\frac{\xi ''}{\xi}+\frac{2\alpha '}{\alpha}\right)  , \notag \\ 
s_{0} & = & -\left[ \frac{\omega^{2}-V}{f'\xi '^{2}}+ \frac{\alpha''}{\alpha}+\left(\frac{f'}{f}+\frac{\xi ''}{\xi}\right)\frac{\alpha '}{\alpha}\right].
\end{eqnarray}

\par
\begin{figure}
\includegraphics[width=13cm]{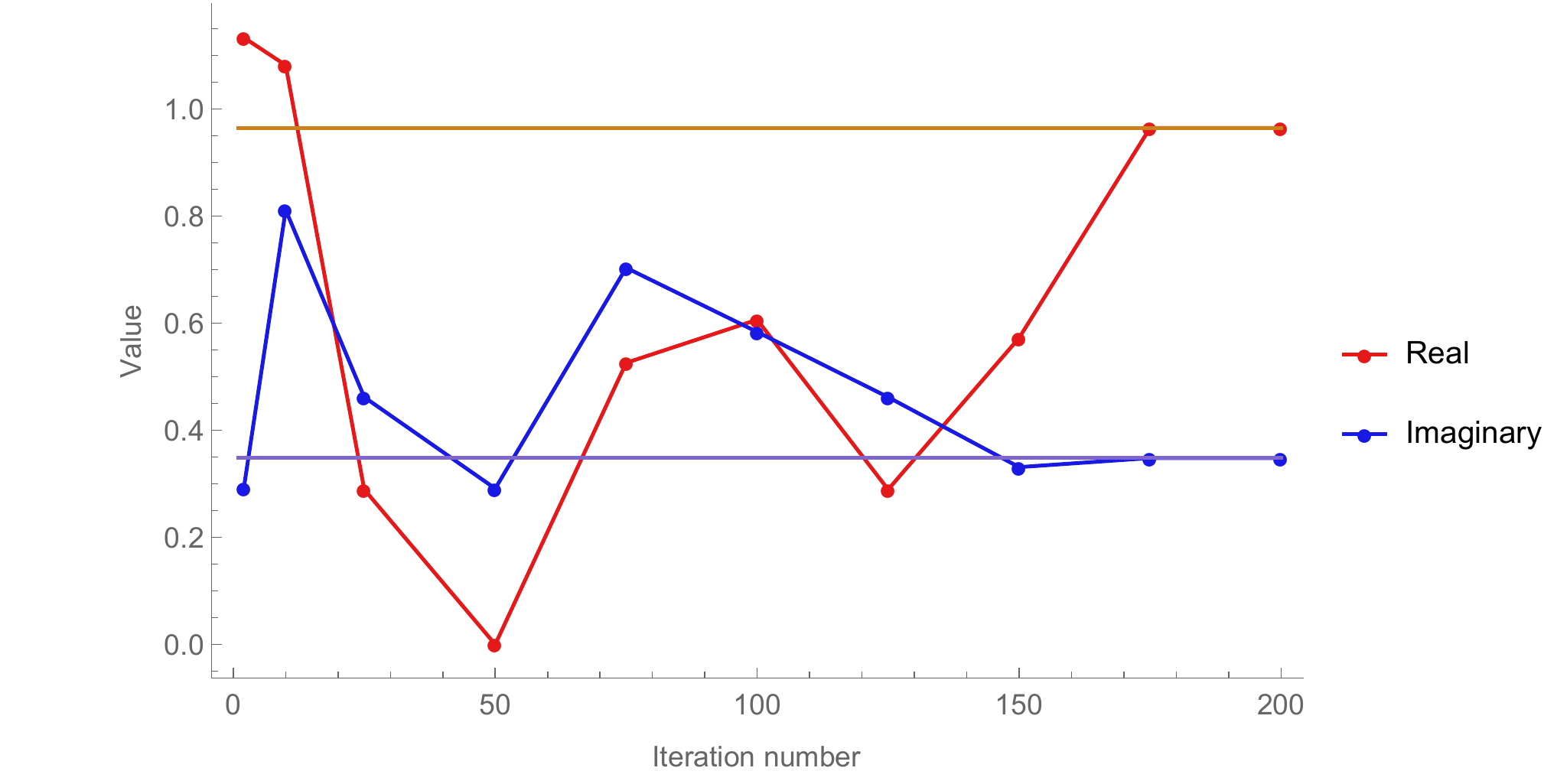}
\caption{\it The relation between iteration number and the value of the real part and the imaginary part of the QNM frequency for $j=3/2$, $Q=1$, and $D=6$.}
\label{QNMD6NL0NTT}
\end{figure}

 We next find the higher order $\lambda$ and $s$ by the relation
\begin{equation}
\lambda_{n}=\lambda_{n-1}'+s_{n-1}+\lambda_{0}\lambda_{n-1} \ \ ; \ \ s_{n}=s_{n-1}'+s_{0}\lambda_{n-1},
\end{equation}
and the corresponding $\omega$ by the equation
\begin{equation}
s_{n}\lambda_{n+1}-s_{n+1}\lambda_{n}=0. \label{lambda s TC}
\end{equation}
Note that
\begin{equation}
f'=\frac{d}{d\xi}f(\xi)\ \ , \ \ \xi '=\left. \frac{d}{dr}\xi(r)\right|_{r=\frac{(r_+)^{1/(D-3)}}{1-\xi^{2}}}\ \ , \ \ \xi''=\frac{d}{d\xi}\xi'\ \ , \ \ \alpha'=\frac{d}{d\xi}\alpha(\xi) .
\end{equation}
Iterating this method for a sufficiently large number of iteration, $\omega$ becomes stable, indicating that this is the QNM we are looking for. For example, for the first mode of the extremal case in 6 dimensions, the relation between iteration number and the QNM frequency is plotted in Fig~\ref{QNMD6NL0NTT} (with $M=1$). This mode is one of the modes that the 3rd and the 6th order WKB methods do not give a reasonable result for, but the AIM does. 

\par
\begin{figure}
\includegraphics[width=13cm]{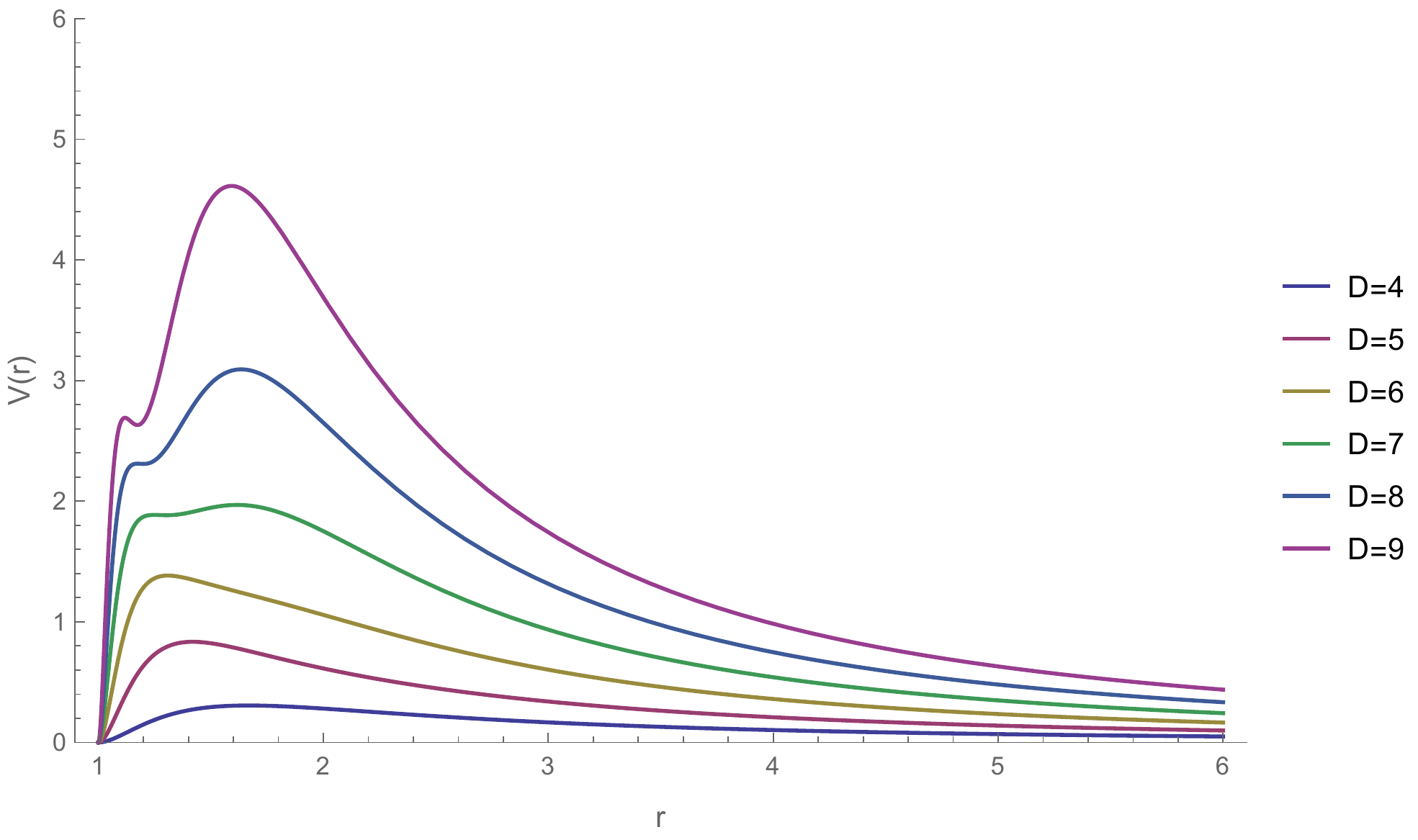}
\caption{\it Potential function with $j=3/2$ and $Q=1$ for $D=4$ to $D=9$.}
\label{Potentiall=0Q=1}
\end{figure}

\begin{figure}
\includegraphics[width=13cm]{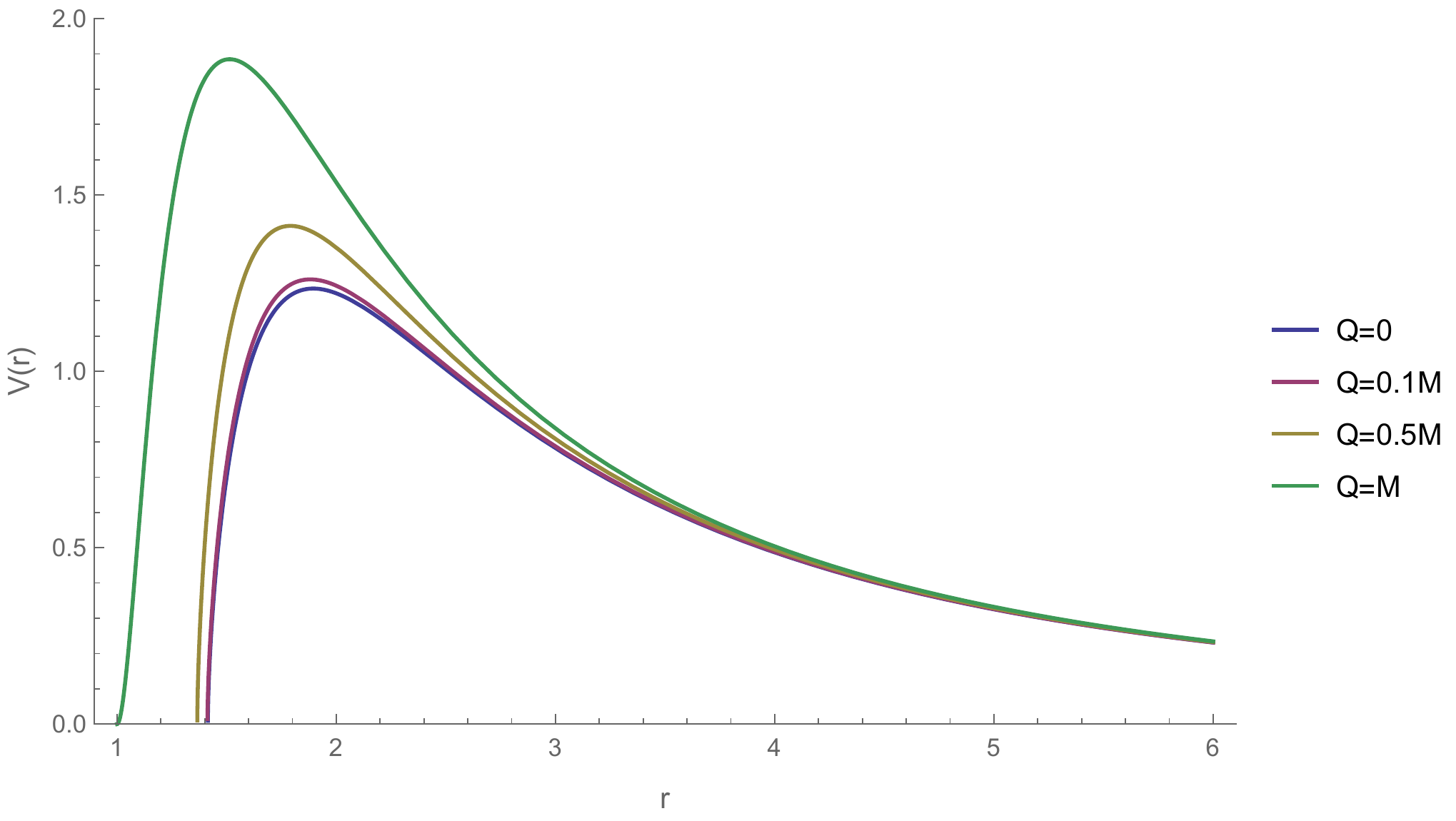}
\caption{\it The non-TT effective potential with $D=5$ and $j=5/2$ for $Q=0$ to $Q = 1$.}
\label{nTTPOTENTIAL}
\end{figure}

 In Tabs. \ref{tab:45DQ=0.1}-\ref{tab:67DQ=M} we present the QNMs for the Reissner-Nordstr\"{o}m black hole for dimensions $D=4$ to $D=7$.  The results are given in units of $M$, that is, we have set $M=1$. We note that as the value of $n$ increases, for fixed values of $l$ ($=j-3/2$) and $D$, the real part of the QNM decreases and the magnitude of the imaginary part increases, this being the same behavior as we have seen for the Schwarzschild black hole. This result suggests that the lower modes are easier to detect compared to the higher less energetic modes. Furthermore, they also decay the slowest. We also note that an increase in the number of dimensions results in the QNM being emitted more energetically. This can be understood by considering the change in the potentials as the dimension is increased as shown in Fig.~\ref{Potentiall=0Q=1}. From $D=4$ to $D=7$ the maximum value of the potential increases as $D$ is increased. Hence, the real part of the QNM frequency would also increase. Lastly, when the charge $Q$ is increased, the real part of the frequency for the same mode increases, while the magnitude of the imaginary part also increases. This is consistent with the change of the effective potentials as $Q$ is increased, as shown in Fig.~\ref{nTTPOTENTIAL}. As $Q$ is increased from 0 to 1 (in units of $M$), the maximum value of the potential increases, hence the real part of the QNM frequency increases. On the other hand, the potential tends to sharpen as $Q$ is increased, this implies that the field can decay more easily, giving a large decay constant, or a large absolute value of the imaginary part of the frequency.

\par Note that in the tables there are several blank entries. The reason for leaving these entries out is that we think the numbers we have obtained are not reliable. For the WKB approximation we found that higher order terms dominate over the lower order terms. This is unreasonable as the WKB method is generated from a series expansion. As for the AIM, the results do not converge when the number of iterations is increased. As such, in these cases we have left these entries as blank.

\par We also find that there is a strong disagreement for the WKB methods in the cases of $Q=M$, for dimensions higher than $7$. The reason for this disagreement is two-fold. The first is again the problem with the WKB series expansion, as mentioned above. 
The second one is due to the peculiar behavior of the effective potential. As shown in Fig. \ref{Potentiall=0Q=1}, it is clear that for the $j=3/2$ potentials in the cases $D>7$, a second local maximum will develop. This happens not just for the $j=3/2$ cases but also for potentals with other $j$ values. For larger values of $j$, the dimension at which the potential will have this behavior is higher. The presence of a second maximum renders the WKB approximation and the AIM unreliable, so we have only listed the results up to $D=7$ in these instances.

\begin{table}[ht]
\caption{\it Low-lying ($n\leq l$, with $l=j-3/2$) spin-3/2 field quasi-normal frequencies using the WKB and the AIM with D = 4,5 and Q=0.1M.}
\centering
\begin{tabular}{| l | l || l | l || l || l | l | l | l || l | l || l | l || l | }
\hline
\multicolumn{5}{|c||}{4 Dimensions} & \multicolumn{5}{c|}{5 Dimensions}\\
\hline
l & n & WKB 3rd order & WKB 6th order & AIM & l & n & WKB 3rd order & WKB 6th order & AIM \\
\hline
0 & 0 &  0.3161-0.0909i &  0.3185-0.0910i &  0.3185-0.0909i & 0 & 0 &  0.6360-0.2147i &  0.6484-0.2191i &  0.6484-0.2191i \\
 \hline
1 & 0 &  0.5370-0.0942i &  0.5375-0.0942i &  0.5374-0.0942i & 1 & 0 &  1.0773-0.2343i &  1.0808-0.2343i &  1.0807-0.2343i \\

1 & 1 &  0.5180-0.2871i &  0.5191-0.2867i &  0.5191-0.2867i & 1 & 1 &  0.9917-0.7241i &  1.0008-0.7209i &  1.0008-0.7209i \\
 \hline

2 & 0 &  0.7423-0.0952i &  0.7425-0.0952i &  0.7424-0.0952i & 2 & 0 &  1.4730-0.2411i &  1.4742-0.2411i &  1.4742-0.2411i \\

2 & 1 &  0.7285-0.2880i &  0.7289-0.2879i &  0.7289-0.2878i & 2 & 1 &  1.4100-0.7347i &  1.4126-0.7342i &  1.4126-0.7342i \\

2 & 2 &  0.7041-0.4860i &  0.7035-0.4871i &  0.7034-0.4870i & 2 & 2 &  1.2998-1.2522i &  1.2922-1.2636i &  1.2922-1.2636i \\
 \hline

3 & 0 &  0.9423-0.0957i &  0.9424-0.0957i &  0.9424-0.0956i & 3 & 0 &  1.8520-0.2443i &  1.8527-0.2443i &  1.8527-0.2443i \\

3 & 1 &  0.9315-0.2884i &  0.9316-0.2884i &  0.9316-0.2883i & 3 & 1 &  1.8018-0.7401i &  1.8040-0.7395i &  1.8040-0.7395i \\

3 & 2 &  0.9114-0.4848i &  0.9109-0.4853i &  0.9109-0.4852i & 3 & 2 &  1.7104-1.2530i &  1.7101-1.2546i &  1.7101-1.2546i \\

3 & 3 &  0.8845-0.6853i &  0.8819-0.6887i &  0.8819-0.6887i & 3 & 3 &  1.5882-1.7838i &  1.5787-1.8012i &  1.5787-1.8012i \\
 \hline

4 & 0 &  1.1398-0.0959i &  1.1399-0.0959i &  1.1398-0.0958i & 4 & 0 &  2.2228-0.2461i &  2.2232-0.2461i &  2.2232-0.2461i \\

4 & 1 &  1.1309-0.2886i &  1.1310-0.2886i &  1.1309-0.2886i & 4 & 1 &  2.1809-0.7432i &  2.1824-0.7429i &  2.1824-0.7429i \\

4 & 2 &  1.1139-0.4840i &  1.1136-0.4842i &  1.1135-0.4842i & 4 & 2 &  2.1027-1.2530i &  2.1031-1.2534i &  2.1031-1.2534i \\

4 & 3 &  1.0905-0.6827i &  1.0886-0.6845i &  1.0886-0.6845i & 4 & 3 &  1.9958-1.7778i &  1.9911-1.7852i &  1.9911-1.7852i \\

4 & 4 &  1.0620-0.8848i &  1.0575-0.8910i &  1.0575-0.8909i & 4 & 4 &  1.8659-2.3161i &  1.8545-2.3415i &  1.8545-2.3415i \\
 \hline

5 & 0 &  1.3360-0.0960i &  1.3360-0.0960i &  1.3360-0.0960i & 5 & 0 &  2.5889-0.2471i &  2.5891-0.2471i &  2.5891-0.2471i \\

5 & 1 &  1.3283-0.2887i &  1.3284-0.2887i &  1.3283-0.2887i & 5 & 1 &  2.5529-0.7451i &  2.5536-0.7450i &  2.5536-0.7450i \\

5 & 2 &  1.3136-0.4834i &  1.3134-0.4836i &  1.3133-0.4835i & 5 & 2 &  2.4846-1.2528i &  2.4838-1.2536i &  2.4838-1.2536i \\

5 & 3 &  1.2929-0.6809i &  1.2916-0.6819i &  1.2916-0.6819i & 5 & 3 &  2.3895-1.7731i &  2.3822-1.7802i &  2.3822-1.7802i \\

5 & 4 &  1.2674-0.8812i &  1.2640-0.8849i &  1.2640-0.8849i & 5 & 4 &  2.2727-2.3056i &  2.2533-2.3315i &  2.2533-2.3315i \\

5 & 5 &  1.2378-1.0842i &  1.2315-1.0935i &  1.2315-1.0935i & 5 & 5 &  2.1371-2.8485i &  2.1024-2.9129i &  2.1024-2.9129i \\
 \hline
\end{tabular}
\label{tab:45DQ=0.1}
\end{table} 

\begin{table}[ht]
\caption{\it Low-lying ($n\leq l$, with $l=j-3/2$) spin-3/2 field quasi-normal frequencies using the WKB and the AIM with D = 6,7 and Q=0.1M}
\centering
\begin{tabular}{| l | l || l | l || l || l | l | l | l || l | l || l | l || l | }
\hline
\multicolumn{5}{|c||}{6 Dimensions} & \multicolumn{5}{c|}{7 Dimensions}\\
\hline
l & n & WKB 3rd order & WKB 6th order & AIM & l & n & WKB 3rd order & WKB 6th order & AIM \\
\hline
0 & 0 &  0.9364-0.3422i &  0.9408-0.3027i &  0.9408-0.3027i & 0 & 0 &  1.2845-0.4857i &  1.2111-0.5136i &  1.2110-0.5136i \\
 \hline

1 & 0 &  1.5164-0.3578i &  1.5292-0.3540i &  1.5292-0.3540i & 1 & 0 &  1.9288-0.4776i &  1.9388-0.4621i &  1.9388-0.4621i \\

1 & 1 &  1.3214-1.1165i &  1.3680-1.0786i &  1.3680-1.0786i & 1 & 1 &  1.5920-1.4970i &  1.6160-1.3961i &  1.6160-1.3961i \\
 \hline

2 & 0 &  2.0379-0.3705i &  2.0418-0.3699i &  2.0418-0.3699i & 2 & 0 &  2.5348-0.4880i &  2.5430-0.4836i &  2.5430-0.4836i \\

2 & 1 &  1.8967-1.1346i &  1.9117-1.1279i &  1.9117-1.1279i & 2 & 1 &  2.2870-1.4974i &  2.3243-1.4620i &  2.3243-1.4620i \\

2 & 2 &  1.6451-1.9511i &  1.6488-1.9502i &  1.6488-1.9502i & 2 & 2 &  1.8288-2.5991i &  1.8401-2.5026i &  1.8401-2.5026i \\
 \hline

3 & 0 &  2.5341-0.3774i &  2.5360-0.3772i &  2.5360-0.3772i & 3 & 0 &  3.1136-0.4968i &  3.1174-0.4954i &  3.1174-0.4954i \\

3 & 1 &  2.4214-1.1470i &  2.4285-1.1444i &  2.4285-1.1444i & 3 & 1 &  2.9167-1.5115i &  2.9352-1.5003i &  2.9352-1.5003i \\

3 & 2 &  2.2133-1.9532i &  2.2124-1.9547i &  2.2124-1.9547i & 3 & 2 &  2.5422-2.5869i &  2.5506-2.5600i &  2.5506-2.5600i \\

3 & 3 &  1.9314-2.8004i &  1.8910-2.8487i &  1.8910-2.8487i & 3 & 3 &  2.0267-3.7434i &  1.9395-3.7554i &  1.9395-3.7554i \\
 \hline

4 & 0 &  3.0173-0.3816i &  3.0183-0.3814i &  3.0183-0.3814i & 4 & 0 &  3.6764-0.5028i &  3.6784-0.5022i &  3.6784-0.5022i \\

4 & 1 &  2.9227-1.1550i &  2.9267-1.1537i &  2.9267-1.1537i & 4 & 1 &  3.5114-1.5233i &  3.5215-1.5186i &  3.5215-1.5186i \\

4 & 2 &  2.7444-1.9555i &  2.7428-1.9567i &  2.7428-1.9567i & 4 & 2 &  3.1931-2.5872i &  3.1959-2.5773i &  3.1959-2.5773i \\

4 & 3 &  2.4979-2.7896i &  2.4678-2.8176i &  2.4678-2.8176i & 4 & 3 &  2.7445-3.7136i &  2.6843-3.7312i &  2.6843-3.7312i \\

4 & 4 &  2.1951-3.6553i &  2.1087-3.7702i &  2.1087-3.7702i & 4 & 4 &  2.1908-4.9064i &  1.9854-5.0614i &  1.9854-5.0614i \\
 \hline

5 & 0 &  3.4926-0.3842i &  3.4932-0.3841i &  3.4932-0.3841i & 5 & 0 &  4.2293-0.5070i &  4.2305-0.5067i &  4.2305-0.5067i \\

5 & 1 &  3.4109-1.1603i &  3.4133-1.1596i &  3.4133-1.1596i & 5 & 1 &  4.0865-1.5321i &  4.0927-1.5296i &  4.0927-1.5296i \\

5 & 2 &  3.2547-1.9572i &  3.2531-1.9581i &  3.2531-1.9581i & 5 & 2 &  3.8085-2.5899i &  3.8092-2.5851i &  3.8092-2.5851i \\

5 & 3 &  3.0353-2.7818i &  3.0130-2.7991i &  3.0130-2.7991i & 5 & 3 &  3.4108-3.6971i &  3.3672-3.7097i &  3.3672-3.7097i \\

5 & 4 &  2.7630-3.6346i &  2.6964-3.7065i &  2.6964-3.7065i & 5 & 4 &  2.9125-4.8608i &  2.7597-4.9595i &  2.7597-4.9595i \\

5 & 5 &  2.4444-4.5127i &  2.3116-4.7073i &  2.3116-4.7073i & 5 & 5 &  2.3296-6.0794i &  1.9987-6.4037i &  1.9987-6.4037i \\
\hline
\end{tabular}
\label{tab:67DQ=0.1M}
\end{table}

\begin{table}[ht]
\caption{\it Low-lying ($n\leq l$, with $l=j-3/2$) spin-3/2 field quasi-normal frequencies using the WKB and the AIM with D = 4,5 and Q=0.5M.}
\centering
\begin{tabular}{| l | l || l | l || l || l | l | l | l || l | l || l | l || l | }
\hline
\multicolumn{5}{|c||}{4 Dimensions} & \multicolumn{5}{c|}{5 Dimensions}\\
\hline
l & n & WKB 3rd order & WKB 6th order & AIM & l & n & WKB 3rd order & WKB 6th order & AIM \\
\hline
0 & 0 &  0.3616-0.0953i &  0.3634-0.0952i &  0.3621-0.0881i & 0 & 0 &  0.7048-0.2213i &  0.8005-0.1499i &  0.7065-0.2259i \\
 \hline

1 & 0 &  0.5905-0.0971i &  0.5908-0.0971i &  0.4050-0.1854i & 1 & 0 &  1.1471-0.2356i &  1.1504-0.2347i &  1.1504-0.2346i \\

1 & 1 &  0.5736-0.2950i &  0.5744-0.2948i &  0.3674-0.3656i & 1 & 1 &  1.0756-0.7253i &  1.0842-0.7160i &  1.0842-0.7159i \\
 \hline

2 & 0 &  0.8044-0.0976i &  0.8046-0.0976i &  0.8045-0.0975i & 2 & 0 &  1.5506-0.2420i &  1.5519-0.2420i &  1.5519-0.2420i \\

2 & 1 &  0.7920-0.2948i &  0.7923-0.2947i &  0.7922-0.2947i & 2 & 1 &  1.4946-0.7364i &  1.4995-0.7345i &  1.4995-0.7345i \\

2 & 2 &  0.7698-0.4967i &  0.7692-0.4977i &  0.7691-0.4976i & 2 & 2 &  1.3974-1.2527i &  1.4031-1.2505i &  1.4031-1.2505i \\
 \hline

3 & 0 &  1.0131-0.0977i &  1.0132-0.0977i &  1.0131-0.0977i & 3 & 0 &  1.9368-0.2448i &  1.9374-0.2448i &  1.9374-0.2448i \\

3 & 1 &  1.0032-0.2945i &  1.0033-0.2945i &  1.0033-0.2944i & 3 & 1 &  1.8913-0.7411i &  1.8931-0.7408i &  1.8931-0.7408i \\

3 & 2 &  0.9849-0.4946i &  0.9844-0.4950i &  0.9844-0.4950i & 3 & 2 &  1.8088-1.2532i &  1.8074-1.2559i &  1.8074-1.2559i \\

3 & 3 &  0.9602-0.6986i &  0.9579-0.7015i &  0.9579-0.7015i & 3 & 3 &  1.6992-1.7820i &  1.6869-1.8027i &  1.6869-1.8027i \\
 \hline

4 & 0 &  1.2193-0.0978i &  1.2193-0.0978i &  1.2193-0.0978i & 4 & 0 &  2.3142-0.2462i &  2.3145-0.2462i &  2.3145-0.2462i \\

4 & 1 &  1.2110-0.2943i &  1.2111-0.2943i &  1.2111-0.2942i & 4 & 1 &  2.2759-0.7433i &  2.2770-0.7432i &  2.2770-0.7432i \\

4 & 2 &  1.1954-0.4932i &  1.1951-0.4934i &  1.1951-0.4934i & 4 & 2 &  2.2047-1.2523i &  2.2036-1.2538i &  2.2036-1.2538i \\

4 & 3 &  1.1738-0.6952i &  1.1722-0.6968i &  1.1722-0.6967i & 4 & 3 &  2.1077-1.7753i &  2.0983-1.7871i &  2.0983-1.7871i \\

4 & 4 &  1.1476-0.9004i &  1.1437-0.9057i &  1.1437-0.9057i & 4 & 4 &  1.9902-2.3108i &  1.9669-2.3518i &  1.9669-2.3518i \\
 \hline

5 & 0 &  1.4240-0.0978i &  1.4241-0.0978i &  1.4241-0.0978i & 5 & 0 &  2.6865-0.2470i &  2.6868-0.2470i &  2.6868-0.2470i \\

5 & 1 &  1.4170-0.2941i &  1.4170-0.2941i &  1.4170-0.2941i & 5 & 1 &  2.6535-0.7445i &  2.6542-0.7444i &  2.6542-0.7444i \\

5 & 2 &  1.4034-0.4922i &  1.4032-0.4923i &  1.4032-0.4923i & 5 & 2 &  2.5910-1.2510i &  2.5901-1.2519i &  2.5901-1.2519i \\

5 & 3 &  1.3843-0.6928i &  1.3831-0.6937i &  1.3831-0.6937i & 5 & 3 &  2.5041-1.7694i &  2.4970-1.7766i &  2.4970-1.7766i \\

5 & 4 &  1.3606-0.8962i &  1.3576-0.8994i &  1.3576-0.8994i & 5 & 4 &  2.3977-2.2993i &  2.3787-2.3250i &  2.3787-2.3250i \\

5 & 5 &  1.3333-1.1022i &  1.3277-1.1101i &  1.3277-1.1101i & 5 & 5 &  2.2747-2.8389i &  2.2403-2.9032i &  2.2403-2.9032i \\
 \hline
\end{tabular}
\label{tab:45DQ=0.5}
\end{table}

\begin{table}[ht]
\caption{\it Low-lying ($n\leq l$, with $l=j-3/2$) spin-3/2 field quasi-normal frequencies using the WKB and the AIM with D = 6,7 and Q=0.5M.}
\centering
\begin{tabular}{| l | l || l | l || l || l | l | l | l || l | l || l | l || l | }
\hline
\multicolumn{5}{|c||}{6 Dimensions} & \multicolumn{5}{c|}{7 Dimensions}\\
\hline
l & n & WKB 3rd order & WKB 6th order & AIM & l & n & WKB 3rd order & WKB 6th order & AIM \\
\hline
0 & 0 &  0.9546-0.1667i &  &  0.9545-0.1666i & 0 & 0 &  1.1968-0.3978i &  &  1.1967-0.3978i \\
 \hline

1 & 0 &  1.5887-0.3502i &  1.5411-0.3743i &  1.5411-0.3743i & 1 & 0 &  1.9675-0.4197i &   &   \\

1 & 1 &  1.4709-1.0935i &  1.2200-1.3800i &  1.2200-1.3800i & 1 & 1 &  1.6232-1.2937i &   &   \\
 \hline

2 & 0 &  2.1178-0.3667i &  2.1190-0.3673i &  2.1190-0.3673i & 2 & 0 &  2.6056-0.4703i &  2.5998-0.4897i &  2.5998-0.4897i \\

2 & 1 &  2.0024-1.1232i &  2.0024-1.1224i &  2.0024-1.1224i & 2 & 1 &  2.4023-1.4443i &  2.3449-1.6313i &  2.3449-1.6313i \\

2 & 2 &  1.8050-1.9277i &  1.7720-1.9392i &  1.7720-1.9392i & 2 & 2 &  2.0392-2.4977i &  1.9354-3.3791i &  1.9354-3.3791i \\
 \hline

3 & 0 &  2.6220-0.3750i &  2.6235-0.3751i &  2.6235-0.3751i & 3 & 0 &  3.1972-0.4877i &  3.1984-0.4909i &  3.1984-0.4909i \\

3 & 1 &  2.5225-1.1395i &  2.5271-1.1384i &  2.5271-1.1384i & 3 & 1 &  3.0274-1.4858i &  3.0248-1.5096i &  3.0248-1.5096i \\

3 & 2 &  2.3417-1.9388i &  2.3351-1.9436i &  2.3351-1.9436i & 3 & 2 &  2.7137-2.5414i &  2.6649-2.6679i &  2.6649-2.6679i \\

3 & 3 &  2.1010-2.7748i &  2.0523-2.8252i &  2.0523-2.8252i & 3 & 3 &  2.2912-3.6638i &  2.1445-4.1160i &  2.1445-4.1160i \\
 \hline

4 & 0 &  3.1119-0.3795i &  3.1129-0.3796i &  3.1129-0.3796i & 4 & 0 &  3.7688-0.4968i &  3.7706-0.4975i &  3.7706-0.4975i \\

4 & 1 &  3.0261-1.1488i &  3.0295-1.1482i &  3.0295-1.1482i & 4 & 1 &  3.6211-1.5062i &  3.6269-1.5094i &  3.6269-1.5094i \\

4 & 2 &  2.8655-1.9440i &  2.8632-1.9468i &  2.8632-1.9468i & 4 & 2 &  3.3407-2.5582i &  3.3302-2.5804i &  3.3302-2.5804i \\

4 & 3 &  2.6457-2.7704i &  2.6165-2.8004i &  2.6165-2.8004i & 4 & 3 &  2.9521-3.6661i &  2.8719-3.7744i &  2.8719-3.7744i \\

4 & 4 &  2.3779-3.6250i &  2.2969-3.7378i &  2.2969-3.7378i & 4 & 4 &  2.4763-4.8284i &  2.2632-5.1748i &  2.2632-5.1748i \\
 \hline

5 & 0 &  3.5931-0.3823i &  3.5937-0.3823i &  3.5937-0.3823i & 5 & 0 &  4.3286-0.5023i &  4.3299-0.5025i &  4.3299-0.5025i \\

5 & 1 &  3.5180-1.1543i &  3.5204-1.1540i &  3.5204-1.1540i & 5 & 1 &  4.1979-1.5186i &  4.2034-1.5187i &  4.2034-1.5187i \\

5 & 2 &  3.3751-1.9464i &  3.3738-1.9481i &  3.3738-1.9481i & 5 & 2 &  3.9462-2.5673i &  3.9440-2.5725i &  3.9440-2.5725i \\

5 & 3 &  3.1756-2.7646i &  3.1555-2.7829i &  3.1555-2.7829i & 5 & 3 &  3.5902-3.6620i &  3.5428-3.7021i &  3.5428-3.7021i \\

5 & 4 &  2.9296-3.6087i &  2.8698-3.6796i &  2.8698-3.6796i & 5 & 4 &  3.1475-4.8060i &  2.9968-4.9618i &  2.9968-4.9618i \\

5 & 5 &  2.6429-4.4752i &  2.5247-4.6616i &  2.5247-4.6616i & 5 & 5 &  2.6308-5.9957i &  2.3192-6.4154i &  2.3192-6.4154i \\
 \hline
\end{tabular}
\label{tab:67DQ=0.5}
\end{table}

\begin{table}[ht]
\caption{\it Low-lying ($n\leq l$, with $l=j-3/2$) spin-3/2 field quasi-normal frequencies using the WKB and the AIM with D = 4,5 and Q=M.}
\centering
\begin{tabular}{| l | l || l | l || l || l | l | l | l || l | l || l | l || l | }
\hline
\multicolumn{5}{|c||}{4 Dimensions} & \multicolumn{5}{c|}{5 Dimensions}\\
\hline
l & n & WKB 3rd order & WKB 6th order & AIM & l & n & WKB 3rd order & WKB 6th order & AIM \\
\hline
0 & 0 &  0.5410-0.0867i &  0.5414-0.0865i & 0.5414-0.0864i & 0 & 0 &  0.8519-0.1875i &   &  \\
\hline

1 & 0 &  0.8173-0.0874i &  0.8174-0.0874i &  0.8174-0.0873i & 1 & 0 &  1.3394-0.2077i &  1.3414-0.2098i &  1.3413-0.2097i \\

1 & 1 &  0.8027-0.2638i &  0.8032-0.2636i &  0.8032-0.2636i & 1 & 1 &  1.2704-0.6320i &  1.2809-0.6422i &  1.2809-0.6422i \\
 \hline

2 & 0 &  1.0810-0.0878i &  1.0811-0.0878i &  1.0811-0.0877i & 2 & 0 &  1.7723-0.2144i &  1.7735-0.2147i &  1.7735-0.2147i \\

2 & 1 &  1.0701-0.2643i &  1.0704-0.2643i &  1.0703-0.2642i & 2 & 1 &  1.7224-0.6490i &  1.7266-0.6494i &  1.7266-0.6494i \\

2 & 2 &  1.0491-0.4435i &  1.0491-0.4435i &  1.0490-0.4435i & 2 & 2 &  1.6288-1.0985i &  1.6337-1.1014i &  1.6337-1.1014i \\
 \hline

3 & 0 &  1.3395-0.0880i &  1.3396-0.0880i &  1.3395-0.0879i & 3 & 0 &  2.1865-0.2174i &  2.1872-0.2175i &  2.1872-0.2175i \\

3 & 1 &  1.3307-0.2646i &  1.3309-0.2646i &  1.3308-0.2645i & 3 & 1 &  2.1461-0.6559i &  2.1486-0.6558i &  2.1486-0.6558i \\

3 & 2 &  1.3136-0.4430i &  1.3136-0.4430i &  1.3135-0.4430i & 3 & 2 &  2.0689-1.1048i &  2.0716-1.1049i &  2.0716-1.1049i \\

3 & 3 &  1.2889-0.6239i &  1.2880-0.6245i &  1.2880-0.6245i & 3 & 3 &  1.9604-1.5675i &  1.9574-1.5732i &  1.9574-1.5732i \\
 \hline

4 & 0 &  1.5953-0.0881i &  1.5953-0.0881i &  1.5953-0.0881i & 4 & 0 &  2.5913-0.2189i &  2.5917-0.2190i &  2.5917-0.2190i \\

4 & 1 &  1.5879-0.2648i &  1.5880-0.2648i &  1.5880-0.2648i & 4 & 1 &  2.5572-0.6595i &  2.5587-0.6594i &  2.5587-0.6594i \\

4 & 2 &  1.5734-0.4427i &  1.5734-0.4427i &  1.5734-0.4427i & 4 & 2 &  2.4912-1.1076i &  2.4929-1.1075i &  2.4929-1.1075i \\

4 & 3 &  1.5523-0.6225i &  1.5517-0.6228i &  1.5517-0.6228i & 4 & 3 &  2.3969-1.5663i &  2.3948-1.5691i &  2.3948-1.5691i \\

4 & 4 &  1.5252-0.8047i &  1.5232-0.8060i &  1.5232-0.8060i & 4 & 4 &  2.2784-2.0370i &  2.2655-2.0507i &  2.2655-2.0507i \\
 \hline

5 & 0 &  1.8494-0.0882i &  1.8494-0.0882i &  1.8494-0.0882i & 5 & 0 &  2.9905-0.2199i &  2.9908-0.2199i &  2.9908-0.2199i \\

5 & 1 &  1.8430-0.2649i &  1.8431-0.2649i &  1.8431-0.2649i & 5 & 1 &  2.9610-0.6616i &  2.9620-0.6616i &  2.9620-0.6616i \\

5 & 2 &  1.8305-0.4425i &  1.8305-0.4425i &  1.8305-0.4425i & 5 & 2 &  2.9034-1.1090i &  2.9045-1.1090i &  2.9045-1.1090i \\

5 & 3 &  1.8121-0.6216i &  1.8117-0.6218i &  1.8117-0.6218i & 5 & 3 &  2.8201-1.5649i &  2.8186-1.5665i &  2.8186-1.5665i \\

5 & 4 &  1.7883-0.8026i &  1.7869-0.8034i &  1.7869-0.8034i & 5 & 4 &  2.7143-2.0307i &  2.7049-2.0388i &  2.7049-2.0388i \\

5 & 5 &  1.7596-0.9857i &  1.7563-0.9879i &  1.7563-0.9879i & 5 & 5 &  2.5887-2.5066i &  2.5646-2.5311i &  2.5646-2.5311i \\
 \hline
\end{tabular}
\label{tab:45DQ=M}
\end{table}

\begin{table}[ht]
\caption{\it Low-lying ($n\leq l$, with $l=j-3/2$) spin-3/2 field quasi-normal frequencies using the WKB and the AIM with D = 6,7 and Q=M.}
\centering
\begin{tabular}{| l | l || l | l || l || l | l | l | l || l | l || l | l || l | }
\hline
\multicolumn{5}{|c||}{6 Dimensions} & \multicolumn{5}{c|}{7 Dimensions}\\
\hline
l & n & WKB 3rd order & WKB 6th order & AIM & l & n & WKB 3rd order & WKB 6th order & AIM \\
\hline
0 & 0 &   &   &  0.9646 - 0.3481i & 0 & 0 &  &  &   \\
 \hline

1 & 0 &  1.7521-0.3082i &   &  1.7516 - 0.3053i & 1 & 0 &   &  &   \\

1 & 1 &  1.6678-0.9419i &  &  1.6970 - 0.9918i & 1 & 1 &  &   &   \\
 \hline

2 & 0 &  2.3058-0.3275i &  2.3115-0.3240i &  2.3081 - 0.3401i & 2 & 0 &  2.7749-0.4333i &  2.8918-0.3419i &  2.7749-0.4333i \\

2 & 1 &  2.2126-0.9992i &  2.2297-0.9708i &  2.2375 - 1.0523i & 2 & 1 &  2.7194-1.3659i &  &  2.7194-1.3659i \\

2 & 2 &  2.0483-1.7097i &  2.0302-1.6426i &  2.1478 - 1.8272i & 2 & 2 &  2.7162-2.4111i &   &  2.7162-2.4111i \\
 \hline

3 & 0 &  2.8322-0.3371i &  2.8345-0.3368i &  2.8345-0.3368i & 3 & 0 &  3.3789-0.4424i &  3.3883-0.4341i &  3.3883-0.4341i \\

3 & 1 &  2.7461-1.0212i &  2.7554-1.0156i &  2.7554-1.0156i & 3 & 1 &  3.2568-1.3524i &  3.3296-1.2462i &  3.3296-1.2462i \\

3 & 2 &  2.5841-1.7318i &  2.5948-1.7106i &  2.5948-1.7106i & 3 & 2 &  3.0467-2.3221i &  &  3.3425-1.7230i \\

3 & 3 &  2.3621-2.4758i &  2.3449-2.4367i &  2.3449-2.4367i & 3 & 3 &  2.7896-3.3500i &  &  3.9401-1.1831i \\
 \hline

4 & 0 &  3.3424-0.3426i &  3.3438-0.3426i &  3.3438-0.3426i & 4 & 0 &  3.9683-0.4525i &  3.9713-0.4516i &  3.9713-0.4516i \\

4 & 1 &  3.2663-1.0344i &  3.2720-1.0332i &  3.2720-1.0332i & 4 & 1 &  3.8448-1.3713i &  3.8600-1.3555i &  3.8600-1.3555i \\

4 & 2 &  3.1198-1.7450i &  3.1269-1.7402i &  3.1269-1.7402i & 4 & 2 &  3.6112-2.3277i &  3.6410-2.2421i &  3.6410-2.2421i \\

4 & 3 &  2.9124-2.4820i &  2.9058-2.4774i &  2.9058-2.4774i & 4 & 3 &  3.2891-3.3340i &  3.3018-3.0389i &  3.3018-3.0389i \\

4 & 4 &  2.6545-3.2475i &  2.6054-3.2637i &  2.6054-3.2637i & 4 & 4 &  2.8991-4.3888i &  2.7814-3.5851i &  2.7814-3.5851i \\
 \hline

5 & 0 &  3.8425-0.3459i &  3.8435-0.3460i &  3.8435-0.3460i & 5 & 0 &  4.5446-0.4595i &  4.5466-0.4595i &  4.5466-0.4595i \\

5 & 1 &  3.7753-1.0426i &  3.7792-1.0422i &  3.7792-1.0422i & 5 & 1 &  4.4302-1.3874i &  4.4389-1.3841i &  4.4389-1.3841i \\

5 & 2 &  3.6441-1.7533i &  3.6494-1.7517i &  3.6494-1.7517i & 5 & 2 &  4.2074-2.3414i &  4.2195-2.3230i &  4.2195-2.3230i \\

5 & 3 &  3.4554-2.4847i &  3.4524-2.4854i &  3.4524-2.4854i & 5 & 3 &  3.8883-3.3348i &  3.8773-3.2798i &  3.8773-3.2798i \\

5 & 4 &  3.2166-3.2404i &  3.1862-3.2573i &  3.1862-3.2573i & 5 & 4 &  3.4880-4.3739i &  3.3887-4.2548i &  3.3887-4.2548i \\

5 & 5 &  2.9346-4.0206i &  2.8505-4.0854i &  2.8505-4.0854i & 5 & 5 &  3.0202-5.4581i &  2.7165-5.2505i &  2.7165-5.2505i \\
 \hline
\end{tabular}
\label{tab:67DQ=M}
\end{table}

\FloatBarrier
\subsection{TT eigenfunctions related}
\FloatBarrier
\par
\begin{figure}
\includegraphics[width=13cm]{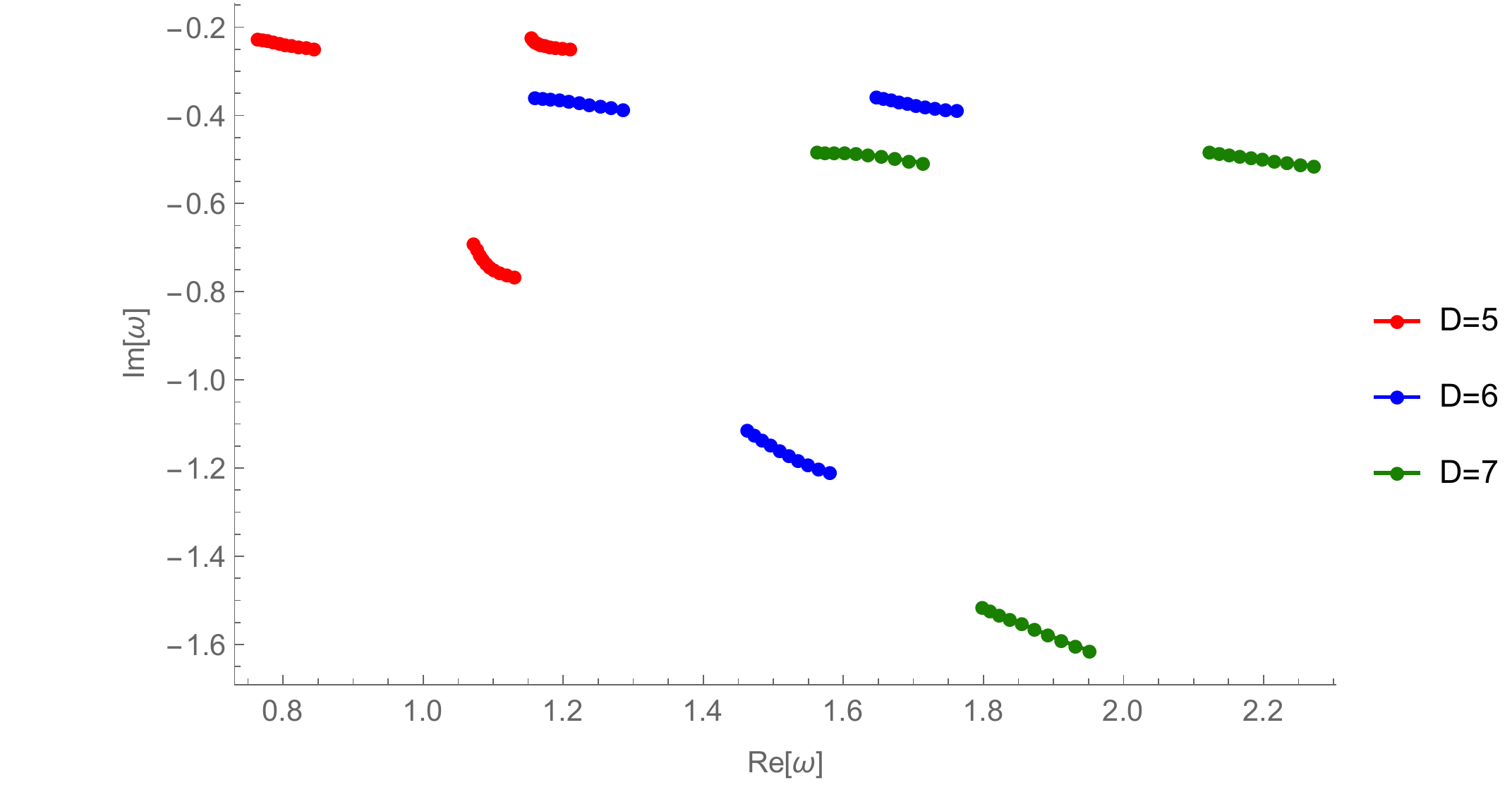}
\caption{\it The change of the first three QNM frequencies when $Q$ increases from 0 to 1.}
\label{RNTTQNMQcomparisionPlot3rd}
\end{figure}

\begin{figure}
\includegraphics[width=13cm]{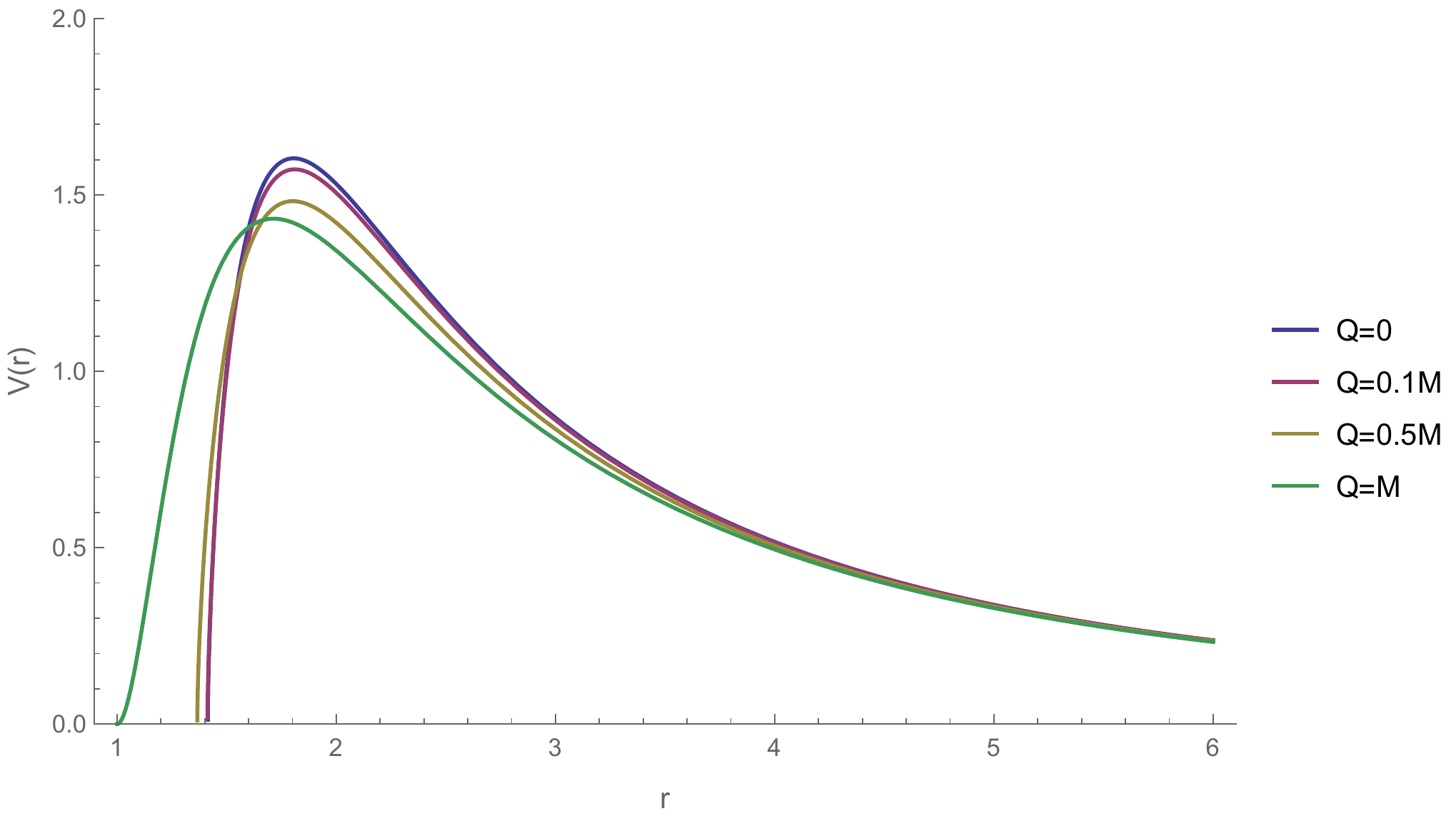}
\caption{\it The TT effective potential of $D=5$ and $j=5/2$ for $Q=0$ to $Q = 1$.}
\label{TTPotential}
\end{figure}

 For the ``TT eigenfunction related" cases, both the WKB and AIM present reasonable results. We have to note that there is no ``TT eigenfunction related" case in the $4$-dimensional Reissner-Nordstr\"{o}m spacetime because of the absence of the TT eigenmodes on the $2$-sphere. In Tabs. \ref{tab:TT45DQ=0.1}-\ref{tab:TT67DQ=1} we present the TT QNMs for $Q=0.1M$, $Q=0.5M$, and $Q=M$ from $D=5$ to $D=7$. The change in QNM frequencies is similar to that for non-TT cases when either $n$ or $D$ is changed. To see how the frequencies are affected by the change in the charge $Q$, we plot in Fig.~\ref{RNTTQNMQcomparisionPlot3rd} the first three modes in more detail for $Q=0$ to 1. The red, blue and green colors represent $5$, $6$, and $7$ dimensions. The result indicates that when $Q$ becomes larger, the real part decreases and the absolute value of imaginary part also decreases. This is consistent with the change of the TT potential with $Q$, which is plotted in Fig.~\ref{TTPotential}. We can see that when $Q$ is increased, the maximum value of the potential decreases. This implies that the real part of the QNM frequency decreases accordingly. In addition to this the potential broadens when $Q$ is increased, so the mode decays slower, which implies that the absolute value of the imaginary part of the frequency becomes smaller. Note that this trend is the exact opposite to that for the non-TT cases for $D<7$.

\begin{table}[ht]
\caption{\it Low-lying ($n\leq l$, with $l=j-3/2$) spin-3/2 field quasi-normal frequencies using the WKB and the AIM with D = 5 and Q=0.1M.}
\centering
\begin{tabular}{| l | l || l | l || l | }
\hline
\multicolumn{5}{|c|}{5 Dimensions}\\
\hline
 l & n & WKB 3rd order & WKB 6th order & AIM \\
\hline
0 & 0 &  0.8443-0.2482i &  0.8538-0.2493i &  0.7755-0.1916i \\
 \hline

1 & 0 &  1.2093-0.2487i &  1.2124-0.2490i &  1.2355-0.2633i \\

1 & 1 &  1.1297-0.7659i &  1.1384-0.7640i &  0.8044-0.6729i \\
 \hline

2 & 0 &  1.5677-0.2490i &  1.5690-0.2490i &  1.5690-0.2490i \\

 2 & 1 &  1.5069-0.7580i &  1.5108-0.7572i &  1.5108-0.7572i \\

2 & 2 &  1.4010-1.2904i &  1.4002-1.2962i &  1.4002-1.2962i \\
 \hline

 3 & 0 &  1.9239-0.2492i &  1.9245-0.2492i &  1.9245-0.2492i \\

3 & 1 &  1.8747-0.7545i &  1.8767-0.7541i &  1.8767-0.7541i \\

 3 & 2 &  1.7854-1.2767i &  1.7840-1.2795i &  1.7840-1.2795i \\

 3 & 3 &  1.6663-1.8167i &  1.6530-1.8392i &  1.6530-1.8392i \\
 \hline

4 & 0 &  2.2791-0.2493i &  2.2795-0.2493i &  2.2795-0.2493i \\

4 & 1 &  2.2379-0.7527i &  2.2390-0.7525i &  2.2390-0.7525i \\

4 & 2 &  2.1608-1.2687i &  2.1596-1.2703i &  2.1596-1.2703i \\

4 & 3 &  2.0557-1.7997i &  2.0451-1.8126i &  2.0451-1.8126i \\

4 & 4 &  1.9280-2.3440i &  1.9017-2.3890i &  1.9017-2.3890i \\
 \hline

5 & 0 &  2.6340-0.2493i &  2.6342-0.2493i &  2.6342-0.2493i \\

5 & 1 &  2.5983-0.7517i &  2.5990-0.7516i &  2.5990-0.7516i \\

5 & 2 &  2.5307-1.2637i &  2.5297-1.2647i &  2.5297-1.2647i \\

5 & 3 &  2.4367-1.7883i &  2.4286-1.7961i &  2.4286-1.7961i \\

5 & 4 &  2.3212-2.3250i &  2.2998-2.3533i &  2.2998-2.3533i \\

5 & 5 &  2.1873-2.8720i &  2.1482-2.9430i &  2.1482-2.9430i \\
 \hline
\end{tabular}
\label{tab:TT45DQ=0.1}
\end{table}

\begin{table}[ht]
\caption{\it Low-lying ($n\leq l$, with $l=j-3/2$) spin-3/2 field quasi-normal frequencies using the WKB and the AIM with D = 6,7 and Q=0.1M.}
\centering
\begin{tabular}{| l | l || l | l || l || l | l | l | l || l | l || l | l || l | }
\hline
\multicolumn{5}{|c||}{6 Dimensions} & \multicolumn{5}{c|}{7 Dimensions}\\
\hline
l & n & WKB 3rd order & WKB 6th order & AIM & l & n & WKB 3rd order & WKB 6th order & AIM \\
\hline
0 & 0 &  1.2853-0.3861i &  1.3064-0.3961i &  1.3063-0.3961i & 0 & 0 &  1.7137-0.5075i &  1.7430-0.5408i &  1.7430-0.5408i \\
 \hline

1 & 0 &  1.7613-0.3889i &  1.7692-0.3917i &  1.7692-0.3917i & 1 & 0 &  2.2717-0.5149i &  2.2844-0.5247i &  2.2844-0.5247i \\

1 & 1 &  1.5809-1.2099i &  1.6097-1.2106i &  1.6097-1.2106i & 1 & 1 &  1.9511-1.6139i &  2.0126-1.6368i &  2.0126-1.6368i \\
 \hline

2 & 0 &  2.2243-0.3901i &  2.2281-0.3906i &  2.2282-0.3904i & 2 & 0 &  2.8099-0.5180i &  2.8176-0.5196i &  2.8176-0.5196i \\

2 & 1 &  2.0867-1.1939i &  2.1008-1.1922i &  2.1008-1.1922i & 2 & 1 &  2.5668-1.5899i &  2.6000-1.5911i &  2.6000-1.5911i \\

2 & 2 &  1.8433-2.0498i &  1.8483-2.0618i &  1.8483-2.0618i & 2 & 2 &  2.1240-2.7539i &  2.1468-2.7766i &  2.1468-2.7766i \\
 \hline

3 & 0 &  2.6826-0.3906i &  2.6846-0.3906i &  2.6846-0.3906i & 3 & 0 &  3.3403-0.5191i &  3.3445-0.5191i &  3.3445-0.5191i \\

3 & 1 &  2.5712-1.1868i &  2.5787-1.1853i &  2.5787-1.1853i & 3 & 1 &  3.1444-1.5797i &  3.1633-1.5766i &  3.1633-1.5766i \\

3 & 2 &  2.3663-2.0196i &  2.3672-2.0240i &  2.3672-2.0240i & 3 & 2 &  2.7746-2.7018i &  2.7858-2.7031i &  2.7858-2.7031i \\

3 & 3 &  2.0899-2.8930i &  2.0549-2.9462i &  2.0549-2.9462i & 3 & 3 &  2.2680-3.9031i &  2.1975-3.9815i &  2.1975-3.9815i \\
 \hline

4 & 0 &  3.1389-0.3909i &  3.1400-0.3908i &  3.1400-0.3908i & 4 & 0 &  3.8673-0.5196i &  3.8696-0.5194i &  3.8696-0.5194i \\

4 & 1 &  3.0451-1.1830i &  3.0493-1.1821i &  3.0493-1.1821i & 4 & 1 &  3.7024-1.5743i &  3.7135-1.5716i &  3.7135-1.5716i \\

4 & 2 &  2.8685-2.0023i &  2.8678-2.0044i &  2.8678-2.0044i & 4 & 2 &  3.3855-2.6730i &  3.3908-2.6705i &  3.3908-2.6705i \\

4 & 3 &  2.6248-2.8550i &  2.5975-2.8847i &  2.5975-2.8847i & 4 & 3 &  2.9404-3.8338i &  2.8874-3.8700i &  2.8874-3.8700i \\

4 & 4 &  2.3261-3.7390i &  2.2457-3.8554i &  2.2457-3.8554i & 4 & 4 &  2.3917-5.0594i &  2.2039-5.2494i &  2.2039-5.2494i \\
 \hline

5 & 0 &  3.5942-0.3910i &  3.5948-0.3910i &  3.5948-0.3910i & 5 & 0 &  4.3926-0.5198i &  4.3940-0.5197i &  4.3940-0.5197i \\

5 & 1 &  3.5129-1.1808i &  3.5156-1.1802i &  3.5156-1.1802i & 5 & 1 &  4.2497-1.5710i &  4.2566-1.5691i &  4.2566-1.5691i \\

5 & 2 &  3.3578-1.9914i &  3.3567-1.9926i &  3.3567-1.9926i & 5 & 2 &  3.9722-2.6553i &  3.9746-2.6529i &  3.9746-2.6529i \\

5 & 3 &  3.1401-2.8297i &  3.1191-2.8476i &  3.1191-2.8476i & 5 & 3 &  3.5760-3.7890i &  3.5363-3.8079i &  3.5363-3.8079i \\

5 & 4 &  2.8703-3.6960i &  2.8067-3.7686i &  2.8067-3.7686i & 5 & 4 &  3.0802-4.9785i &  2.9360-5.0898i &  2.9360-5.0898i \\

5 & 5 &  2.5549-4.5872i &  2.4278-4.7818i &  2.4278-4.7818i & 5 & 5 &  2.5001-6.2217i &  2.1857-6.5663i &  2.1857-6.5663i \\
 \hline
\end{tabular}
\label{tab:TT67DQ=0.1}
\end{table}

\begin{table}[ht]
\caption{\it Low-lying ($n\leq l$, with $l=j-3/2$) spin-3/2 field quasi-normal frequencies using the WKB and the AIM with D = 5 and Q=0.5M.}
\centering
\begin{tabular}{| l | l || l | l || l | }
\hline
\multicolumn{5}{|c|}{5 Dimensions}\\
\hline
 l & n & WKB 3rd order & WKB 6th order & AIM \\
\hline
0 & 0 &  0.8029-0.2388i &  0.8131-0.2411i &  0.3915-0.2511i \\
 \hline

1 & 0 &  1.1737-0.2414i &  1.1769-0.2418i &  1.1768-0.2418i \\

1 & 1 &  1.0949-0.7434i &  1.1046-0.7419i &  1.1045-0.7419i \\
 \hline

 2 & 0 &  1.5374-0.2427i &  1.5387-0.2428i &  1.5387-0.2428i \\

2 & 1 &  1.4783-0.7388i &  1.4825-0.7381i &  1.4825-0.7381i \\

2 & 2 &  1.3751-1.2575i &  1.3754-1.2628i &  1.3754-1.2628i \\
 \hline

3 & 0 &  1.8990-0.2436i &  1.8996-0.2436i &  1.8996-0.2436i \\

3 & 1 &  1.8516-0.7375i &  1.8537-0.7372i &  1.8537-0.7372i \\

3 & 2 &  1.7654-1.2475i &  1.7647-1.2502i &  1.7647-1.2502i \\

3 & 3 &  1.6504-1.7747i &  1.6389-1.7957i &  1.6389-1.7957i \\
 \hline

4 & 0 &  2.2597-0.2442i &  2.2601-0.2442i &  2.2601-0.2442i \\

4 & 1 &  2.2202-0.7372i &  2.2213-0.7370i &  2.2213-0.7370i \\

4 & 2 &  2.1463-1.2422i &  2.1454-1.2437i &  2.1454-1.2437i \\

4 & 3 &  2.0453-1.7616i &  2.0361-1.7736i &  2.0361-1.7736i \\

4 & 4 &  1.9228-2.2938i &  1.8992-2.3356i &  1.8992-2.3356i \\
 \hline

5 & 0 &  2.6201-0.2446i &  2.6203-0.2446i &  2.6203-0.2446i \\

5 & 1 &  2.5860-0.7373i &  2.5867-0.7372i &  2.5867-0.7372i \\

5 & 2 &  2.5214-1.2392i &  2.5207-1.2400i &  2.5207-1.2400i \\

5 & 3 &  2.4316-1.7530i &  2.4245-1.7603i &  2.4245-1.7603i \\

5 & 4 &  2.3212-2.2786i &  2.3020-2.3048i &  2.3020-2.3048i \\

5 & 5 &  2.1933-2.8141i &  2.1580-2.8799i &  2.1580-2.8799i \\
 \hline
\end{tabular}
\label{tab:TT45DQ=0.5}
\end{table}

\begin{table}[ht]
\caption{\it Low-lying ($n\leq l$, with $l=j-3/2$) spin-3/2 field quasi-normal frequencies using the WKB and the AIM with D = 6,7 and Q=0.5M.}
\centering
\begin{tabular}{| l | l || l | l || l || l | l | l | l || l | l || l | l || l | }
\hline
\multicolumn{5}{|c||}{6 Dimensions} & \multicolumn{5}{c|}{7 Dimensions}\\
\hline
l & n & WKB 3rd order & WKB 6th order & AIM & l & n & WKB 3rd order & WKB 6th order & AIM \\
\hline
0 & 0 &  1.2225-0.3706i &  1.2484-0.3817i &  1.2484-0.3816i & 0 & 0 &  1.6352-0.4888i &  1.6769-0.5175i &  1.6769-0.5175i \\
 \hline

1 & 0 &  1.7034-0.3764i &  1.7124-0.3790i &  1.7130-0.3777i & 1 & 0 &  2.1983-0.4993i &  2.2147-0.5067i &  2.2147-0.5067i \\

1 & 1 &  1.5222-1.1710i &  1.5568-1.1692i &  1.5568-1.1692i & 1 & 1 &  1.8724-1.5650i &  1.9497-1.5713i &  1.9497-1.5713i \\
 \hline

2 & 0 &  2.1698-0.3789i &  2.1739-0.3794i &  2.1739-0.3794i & 2 & 0 &  2.7387-0.5037i &  2.7474-0.5046i &  2.7471-0.5046i \\

2 & 1 &  2.0344-1.1593i &  2.0504-1.1572i &  2.0504-1.1572i & 2 & 1 &  2.4972-1.5452i &  2.5363-1.5413i &  2.5363-1.5413i \\

2 & 2 &  1.7938-1.9903i &  1.8052-1.9976i &  1.8052-1.9976i & 2 & 2 &  2.0540-2.6772i &  2.0923-2.6761i &  2.0923-2.6761i \\
 \hline

3 & 0 &  2.6314-0.3803i &  2.6335-0.3804i &  2.6335-0.3804i & 3 & 0 &  3.2711-0.5057i &  3.2757-0.5056i &  3.2757-0.5056i \\

3 & 1 &  2.5230-1.1553i &  2.5313-1.1538i &  2.5313-1.1538i & 3 & 1 &  3.0786-1.5383i &  3.1001-1.5339i &  3.1001-1.5339i \\

3 & 2 &  2.3230-1.9653i &  2.3273-1.9684i &  2.3273-1.9684i & 3 & 2 &  2.7136-2.6303i &  2.7337-2.6236i &  2.7337-2.6236i \\

3 & 3 &  2.0525-2.8150i &  2.0259-2.8605i &  2.0259-2.8605i & 3 & 3 &  2.2117-3.8007i &  2.1600-3.8499i &  2.1600-3.8499i \\
 \hline

4 & 0 &  3.0911-0.3813i &  3.0923-0.3813i &  3.0923-0.3813i & 4 & 0 &  3.8003-0.5069i &  3.8028-0.5067i &  3.8028-0.5067i \\

4 & 1 &  3.0003-1.1538i &  3.0051-1.1529i &  3.0051-1.1529i & 4 & 1 &  3.6393-1.5354i &  3.6518-1.5324i &  3.6518-1.5324i \\

4 & 2 &  2.8293-1.9521i &  2.8307-1.9537i &  2.8307-1.9537i & 4 & 2 &  3.3291-2.6061i &  3.3398-2.6008i &  3.3398-2.6008i \\

4 & 3 &  2.5928-2.7828i &  2.5712-2.8088i &  2.5712-2.8088i & 4 & 3 &  2.8920-3.7372i &  2.8530-3.7610i &  2.8530-3.7610i \\

4 & 4 &  2.3025-3.6437i &  2.2336-3.7479i &  2.2336-3.7479i & 4 & 4 &  2.3523-4.9324i &  2.1898-5.0856i &  2.1898-5.0856i \\
 \hline

5 & 0 &  3.5499-0.3819i &  3.5506-0.3819i &  3.5506-0.3819i & 5 & 0 &  4.3278-0.5077i &  4.3294-0.5076i &  4.3294-0.5076i \\

5 & 1 &  3.4716-1.1532i &  3.4746-1.1527i &  3.4746-1.1527i & 5 & 1 &  4.1890-1.5341i &  4.1967-1.5323i &  4.1967-1.5323i \\

5 & 2 &  3.3221-1.9444i &  3.3224-1.9454i &  3.3224-1.9454i & 5 & 2 &  3.9188-2.5922i &  3.9248-2.5888i &  3.9248-2.5888i \\

5 & 3 &  3.1120-2.7621i &  3.0951-2.7781i &  3.0951-2.7781i & 5 & 3 &  3.5322-3.6979i &  3.5026-3.7110i &  3.5026-3.7110i \\

5 & 4 &  2.8514-3.6067i &  2.7965-3.6724i &  2.7965-3.6724i & 5 & 4 &  3.0476-4.8579i &  2.9241-4.9499i &  2.9241-4.9499i \\

5 & 5 &  2.5466-4.4755i &  2.4344-4.6523i &  2.4344-4.6523i & 5 & 5 &  2.4801-6.0708i &  2.1989-6.3684i &  2.1989-6.3684i \\
 \hline
\end{tabular}
\label{tab:TT67DQ=0.5}
\end{table}

\begin{table}[ht]
\caption{\it Low-lying ($n\leq l$, with $l=j-3/2$) spin-3/2 field quasi-normal frequencies using the WKB and the AIM with D = 5 and Q=M.}
\centering
\begin{tabular}{| l | l || l | l || l | }
\hline
\multicolumn{5}{|c|}{5 Dimensions}\\
\hline
 l & n & WKB 3rd order & WKB 6th order & AIM \\
\hline
 0 & 0 &  0.7636-0.2260i &  0.7748-0.2267i &  0.7748-0.2266i \\
 \hline
 1 & 0 &  1.1542-0.2239i &  1.1579-0.2242i &  1.1578-0.2242i \\

 1 & 1 &  1.0713-0.6900i &  1.0828-0.6866i &  1.0827-0.6866i \\
 \hline

 2 & 0 &  1.5404-0.2232i &  1.5420-0.2233i &  1.5420-0.2233i \\

 2 & 1 &  1.4797-0.6787i &  1.4853-0.6776i &  1.4853-0.6776i \\

 2 & 2 &  1.3701-1.1560i &  1.3734-1.1563i &  1.3734-1.1563i \\
 \hline

 3 & 0 &  1.9256-0.2229i &  1.9264-0.2229i &  1.9264-0.2229i \\

3 & 1 &  1.8779-0.6740i &  1.8809-0.6735i &  1.8809-0.6735i \\

3 & 2 &  1.7885-1.1394i &  1.7906-1.1393i &  1.7906-1.1393i \\

3 & 3 &  1.6656-1.6221i &  1.6574-1.6323i &  1.6574-1.6323i \\
 \hline

4 & 0 &  2.3105-0.2227i &  2.3110-0.2227i &  2.3110-0.2227i \\

4 & 1 &  2.2713-0.6717i &  2.2731-0.6714i &  2.2731-0.6714i \\

4 & 2 &  2.1961-1.1305i &  2.1975-1.1304i &  2.1975-1.1304i \\

4 & 3 &  2.0903-1.6027i &  2.0852-1.6076i &  2.0852-1.6076i \\

4 & 4 &  1.9591-2.0886i &  1.9383-2.1120i &  1.9383-2.1120i \\
 \hline

5 & 0 &  2.6953-0.2226i &  2.6957-0.2226i &  2.6957-0.2226i \\

5 & 1 &  2.6620-0.6703i &  2.6631-0.6701i &  2.6631-0.6701i \\

5 & 2 &  2.5973-1.1252i &  2.5982-1.1252i &  2.5982-1.1252i \\

5 & 3 &  2.5048-1.5905i &  2.5014-1.5932i &  2.5014-1.5932i \\

5 & 4 &  2.3884-2.0674i &  2.3738-2.0806i &  2.3738-2.0806i \\

5 & 5 &  2.2513-2.5556i &  2.2176-2.5942i &  2.2176-2.5942i \\
 \hline
\end{tabular}
\label{tab:TT45DQ=1}
\end{table}

\begin{table}[ht]
\caption{\it Low-lying ($n\leq l$, with $l=j-3/2$) spin-3/2 field quasi-normal frequencies using the WKB and the AIM with D = 6,7 and Q=M.}
\centering
\begin{tabular}{| l | l || l | l || l || l | l | l | l || l | l || l | l || l | }
\hline
\multicolumn{5}{|c||}{6 Dimensions} & \multicolumn{5}{c|}{7 Dimensions}\\
\hline
l & n & WKB 3rd order & WKB 6th order & AIM & l & n & WKB 3rd order & WKB 6th order & AIM \\
\hline
0 & 0 &  1.1597-0.3598i &  1.1816-0.3675i &  1.1816-0.3674i & 0 & 0 &  1.5619-0.4825i &  1.5886-0.5104i &  1.5886-0.5104i \\
 \hline

1 & 0 &  1.6468-0.3575i &  1.6557-0.3593i &  1.6557-0.3593i & 1 & 0 &  2.1226-0.4825i &  2.1369-0.4882i &  2.1369-0.4882i \\

1 & 1 &  1.4621-1.1141i &  1.4941-1.1081i &  1.4941-1.1081i & 1 & 1 &  1.7977-1.5148i &  1.8602-1.5197i &  1.8602-1.5197i \\
 \hline

2 & 0 &  2.1237-0.3565i &  2.1282-0.3568i &  2.1282-0.3568i & 2 & 0 &  2.6667-0.4819i &  2.6752-0.4824i &  2.6752-0.4824i \\

2 & 1 &  1.9865-1.0905i &  2.0037-1.0867i &  2.0037-1.0867i & 2 & 1 &  2.4248-1.4788i &  2.4613-1.4728i &  2.4613-1.4728i \\

2 & 2 &  1.7385-1.8755i &  1.7499-1.8733i &  1.7499-1.8733i & 2 & 2 &  1.9778-2.5675i &  2.0051-2.5593i &  2.0051-2.5593i \\
 \hline

3 & 0 &  2.5979-0.3559i &  2.6003-0.3559i &  2.6003-0.3559i & 3 & 0 &  3.2055-0.4811i &  3.2103-0.4809i &  3.2103-0.4809i \\

3 & 1 &  2.4891-1.0802i &  2.4990-1.0780i &  2.4990-1.0780i & 3 & 1 &  3.0132-1.4630i &  3.0353-1.4574i &  3.0353-1.4574i \\

3 & 2 &  2.2845-1.8379i &  2.2927-1.8346i &  2.2927-1.8346i & 3 & 2 &  2.6446-2.5031i &  2.6659-2.4884i &  2.6659-2.4884i \\

3 & 3 &  2.0038-2.6373i &  1.9791-2.6610i &  1.9791-2.6610i & 3 & 3 &  2.1348-3.6254i &  2.0774-3.6483i &  2.0774-3.6483i \\
 \hline

4 & 0 &  3.0710-0.3555i &  3.0725-0.3555i &  3.0725-0.3555i & 4 & 0 &  3.7422-0.4805i &  3.7452-0.4803i &  3.7452-0.4803i \\

4 & 1 &  2.9808-1.0748i &  2.9869-1.0736i &  2.9869-1.0736i & 4 & 1 &  3.5823-1.4546i &  3.5963-1.4508i &  3.5963-1.4508i \\

4 & 2 &  2.8077-1.8174i &  2.8134-1.8150i &  2.8134-1.8150i & 4 & 2 &  3.2705-2.4682i &  3.2860-2.4568i &  3.2860-2.4568i \\

4 & 3 &  2.5640-2.5920i &  2.5492-2.6021i &  2.5492-2.6021i & 4 & 3 &  2.8265-3.5433i &  2.7943-3.5431i &  2.7943-3.5431i \\

4 & 4 &  2.2621-3.4002i &  2.1950-3.4647i &  2.1950-3.4647i & 4 & 4 &  2.2767-4.6878i &  2.1118-4.7819i &  2.1118-4.7819i \\
 \hline

5 & 0 &  3.5438-0.3552i &  3.5448-0.3552i &  3.5448-0.3552i & 5 & 0 &  4.2782-0.4800i &  4.2801-0.4799i &  4.2801-0.4799i \\

5 & 1 &  3.4667-1.0716i &  3.4707-1.0709i &  3.4707-1.0709i & 5 & 1 &  4.1410-1.4495i &  4.1502-1.4471i &  4.1502-1.4471i \\

5 & 2 &  3.3169-1.8051i &  3.3209-1.8036i &  3.3209-1.8036i & 5 & 2 &  3.8711-2.4475i &  3.8824-2.4397i &  3.8824-2.4397i \\

5 & 3 &  3.1025-2.5636i &  3.0932-2.5684i &  3.0932-2.5684i & 5 & 3 &  3.4802-3.4920i &  3.4614-3.4866i &  3.4614-3.4866i \\

5 & 4 &  2.8329-3.3502i &  2.7864-3.3853i &  2.7864-3.3853i & 5 & 4 &  2.9864-4.5939i &  2.8744-4.6353i &  2.8744-4.6353i \\

5 & 5 &  2.5157-4.1645i &  2.4036-4.2794i &  2.4036-4.2794i & 5 & 5 &  2.4079-5.7543i &  2.1250-5.9489i &  2.1250-5.9489i \\
 \hline
\end{tabular}
\label{tab:TT67DQ=1}
\end{table}

\FloatBarrier
\section{Absorption probabilities}\label{Sec:Abs}

\par In this section we present the absorption probabilities associated with our spin-3/2 fields near a Reissner-Nordstr\"{o}m black hole. We use the same approach as in Ref.\cite{Chen2016}, and as such only present the analysis of the results here, and refer the reader to Ref. \cite{Chen2016} for the implementation of the method.

\FloatBarrier
\subsection{Non-TT eigenmodes related}
\FloatBarrier
\par
\begin{figure}
\begin{subfigure}{0.48\textwidth}
\includegraphics[width=\textwidth]{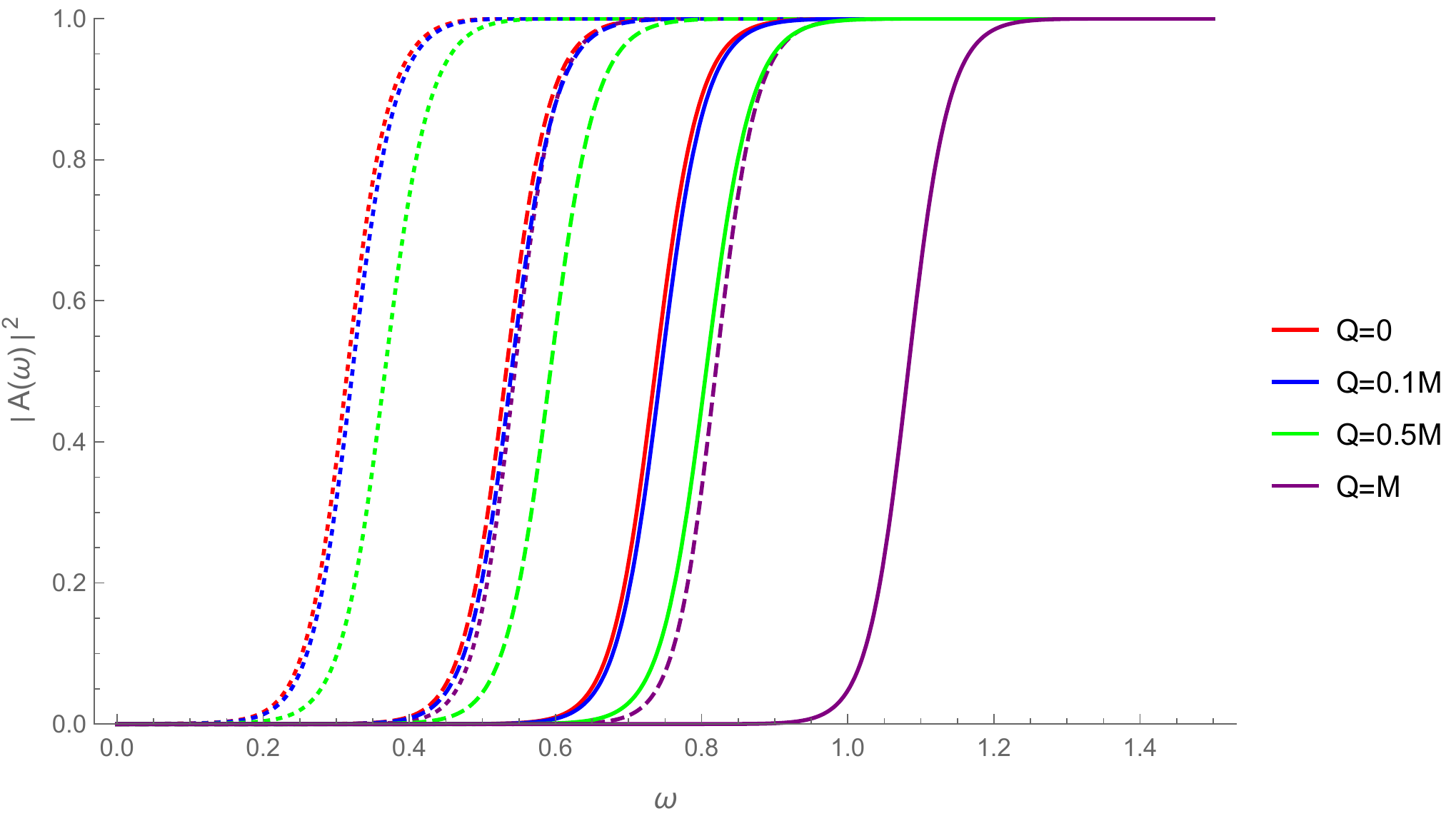}
\caption{$D=4$}
\label{Fig:AP15}
\end{subfigure}
\hfill
\begin{subfigure}{0.48\textwidth}
\includegraphics[width=\textwidth]{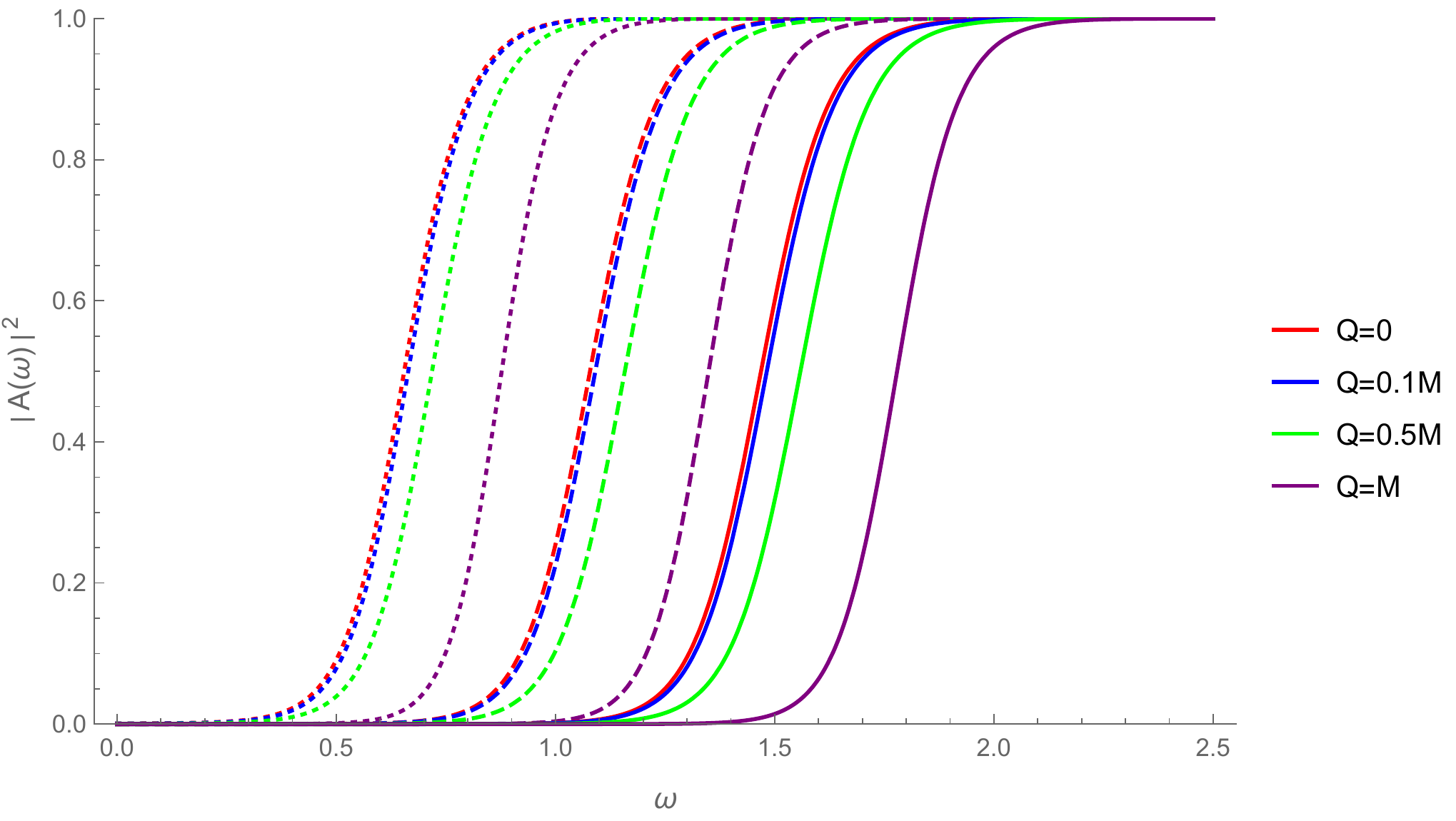}
\caption{$D=5$}
\label{Fig:AP25}
\end{subfigure}
\begin{subfigure}{0.48\textwidth}
\includegraphics[width=\textwidth]{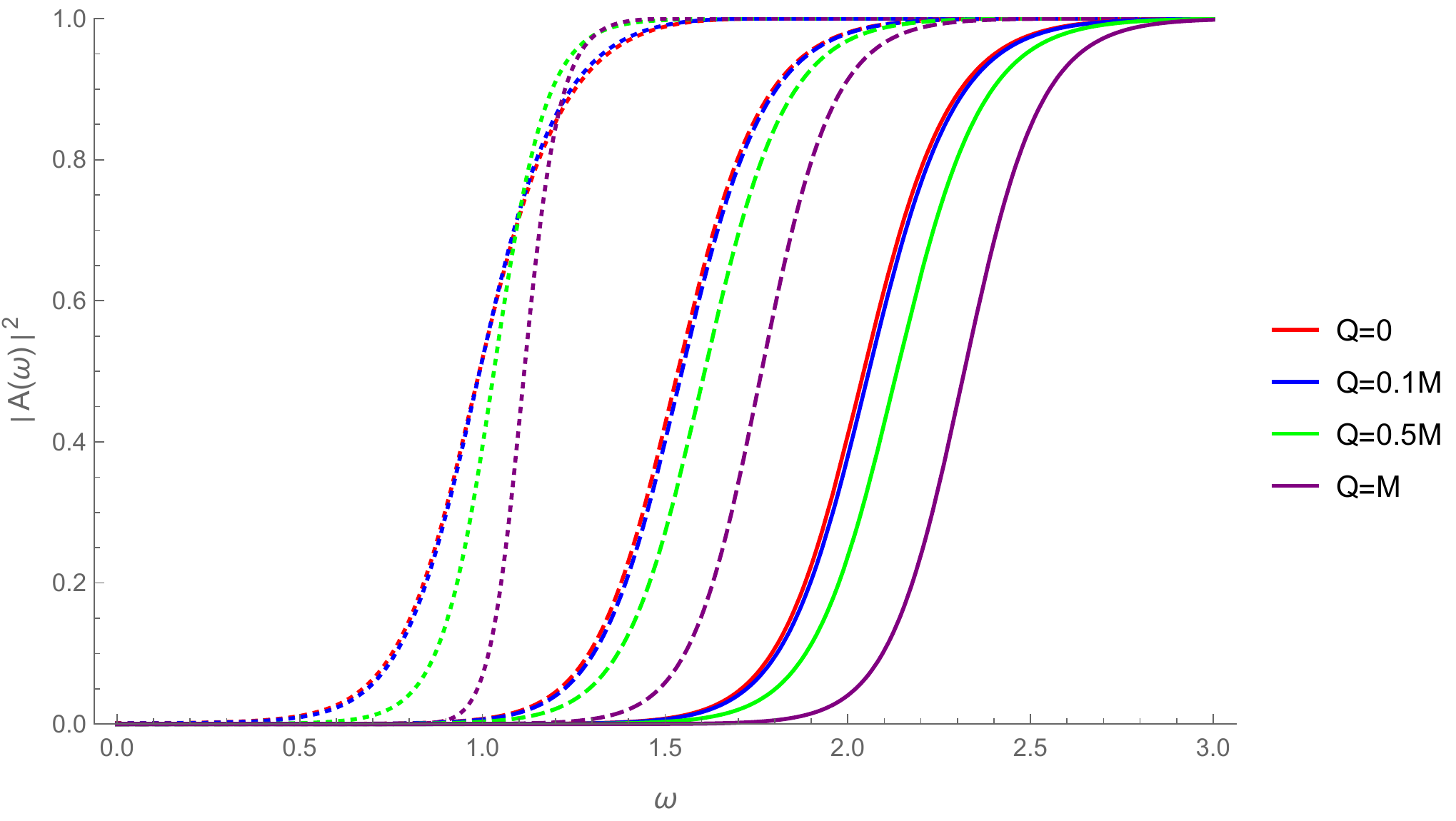}
\caption{$D=6$}
\label{Fig:AP25}
\end{subfigure}
\begin{subfigure}{0.48\textwidth}
\includegraphics[width=\textwidth]{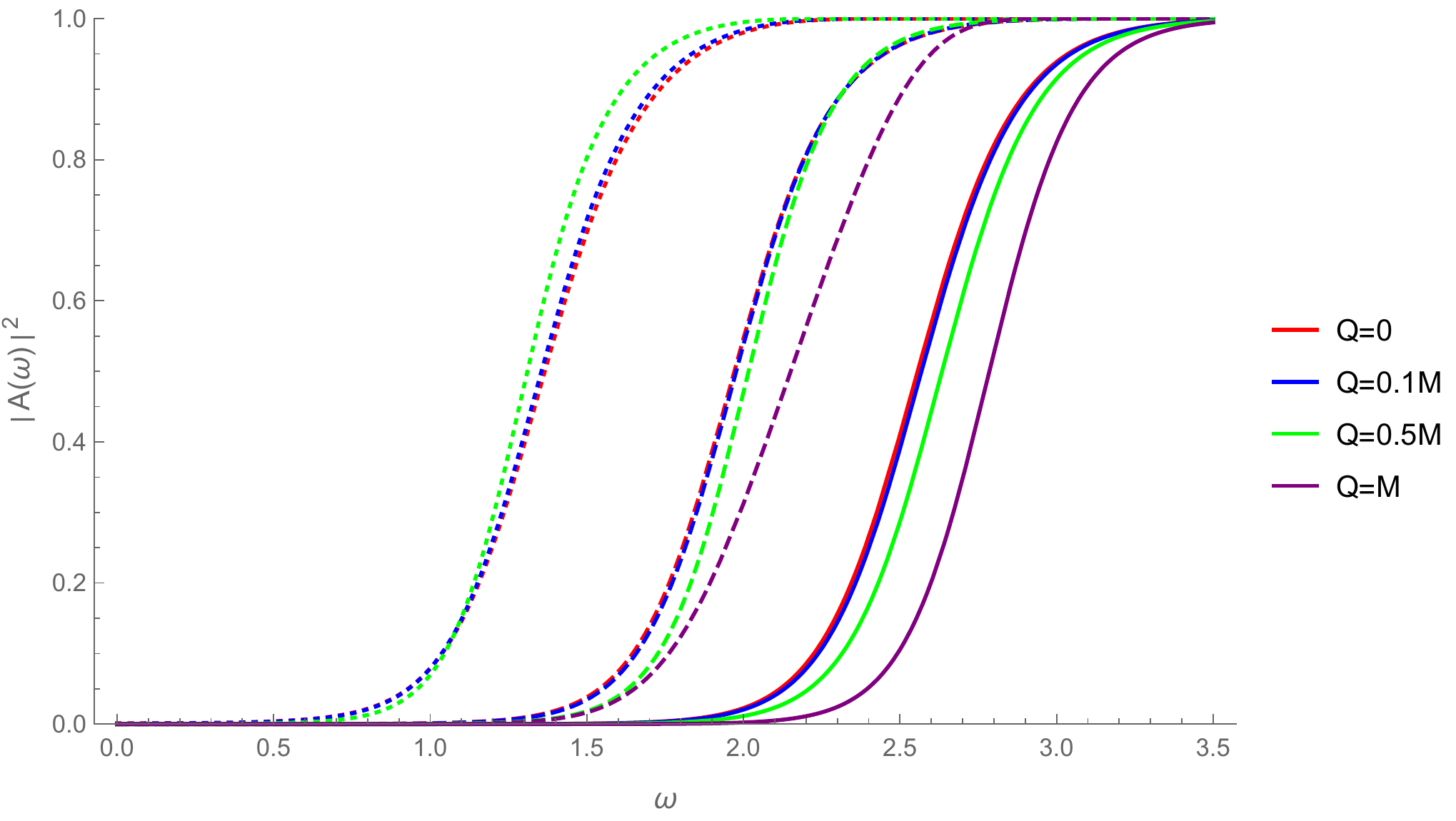}
\caption{$D=7$}
\label{Fig:AP25}
\end{subfigure}
\caption{\it Spin-3/2 field absorption probabilities for various dimensions and $j=3/2$ (Dotted), $j=5/2$ (Dashed) and $j=7/2$ (Solid).}
\label{Fig:APHD}
\end{figure}

 In Fig.~\ref{Fig:APHD} we see that for a specific $Q$ and $D$, the behavior of the absorption probability shifts from lower energy to higher energy as $j$ increases, and this trend is similar to the Schwarzschild case. For fixed $j$ and $Q$, we can compare the scale of each subplot and realize a lower energy to higher energy shift as $D$ increase. For a fixed $j$ and $D$, the absorption probability also shifts from left to right as $Q$ increases. This is due to the maximum value of the corresponding potential increasing as $Q$ increases, as shown in Fig.~\ref{nTTPOTENTIAL}  for the case of $D=5$, $j=5/2$. An exception is in $j=3/2$, $D=7$ case, where the curve shifts to the left instead. This is because the maximum value of the potential decreases instead of increasing as $Q$ is increased. Moreover, we have left out the absorption probability in the case of $j=3/2$, $D=7$, and $Q=1$. We could not obtain a satisfactory curve for this case and believe this is due to the fact that the effective potential has two local maxima rather than one, thus rendering the WKB approximation inapplicable. 

\FloatBarrier
\subsection{TT eigenmodes related}
\begin{figure}
\begin{subfigure}{0.48\textwidth}
\includegraphics[width=\textwidth]{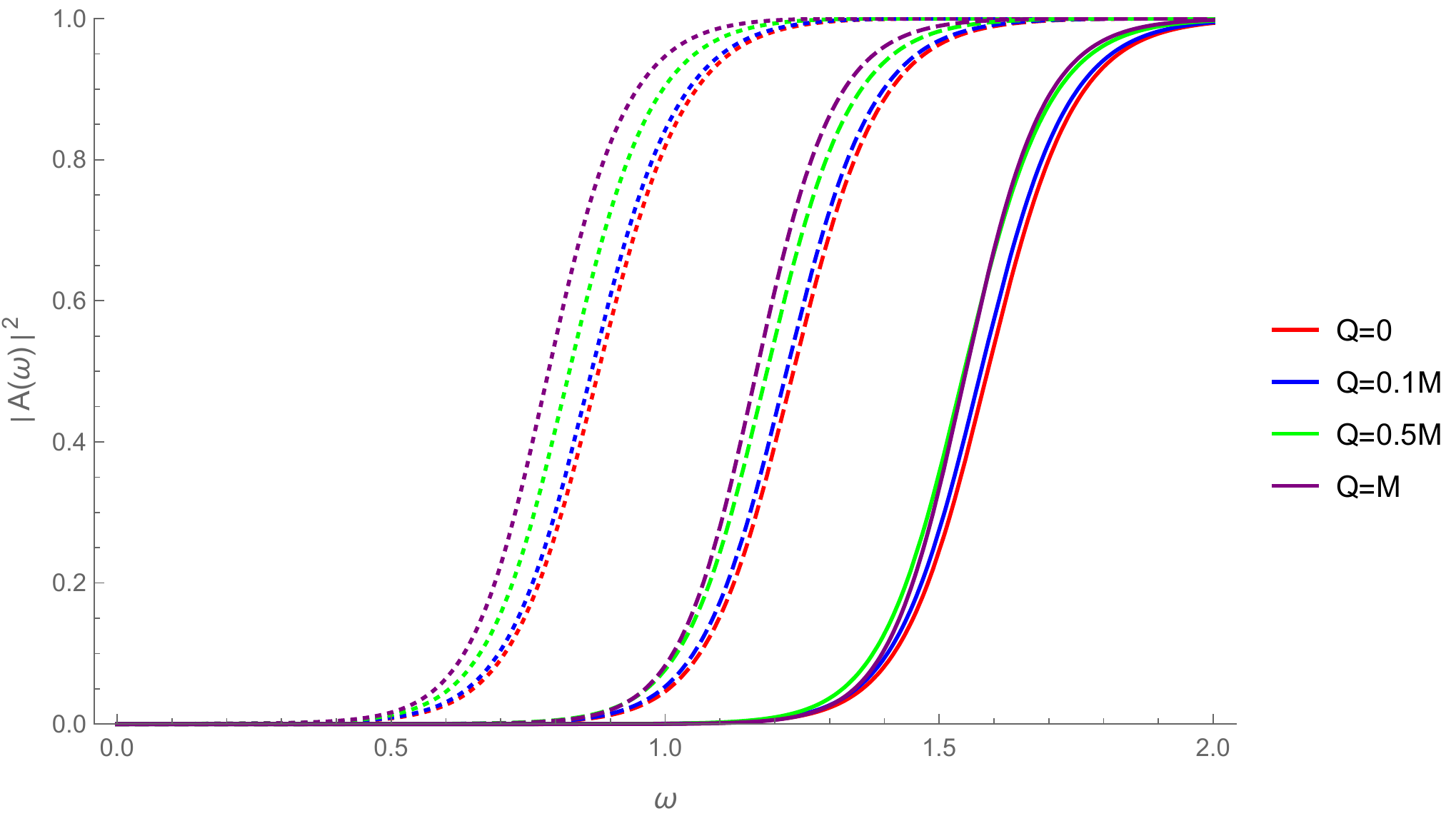}
\caption{$D=5$}
\label{Fig:AP25}
\end{subfigure}
\begin{subfigure}{0.48\textwidth}
\includegraphics[width=\textwidth]{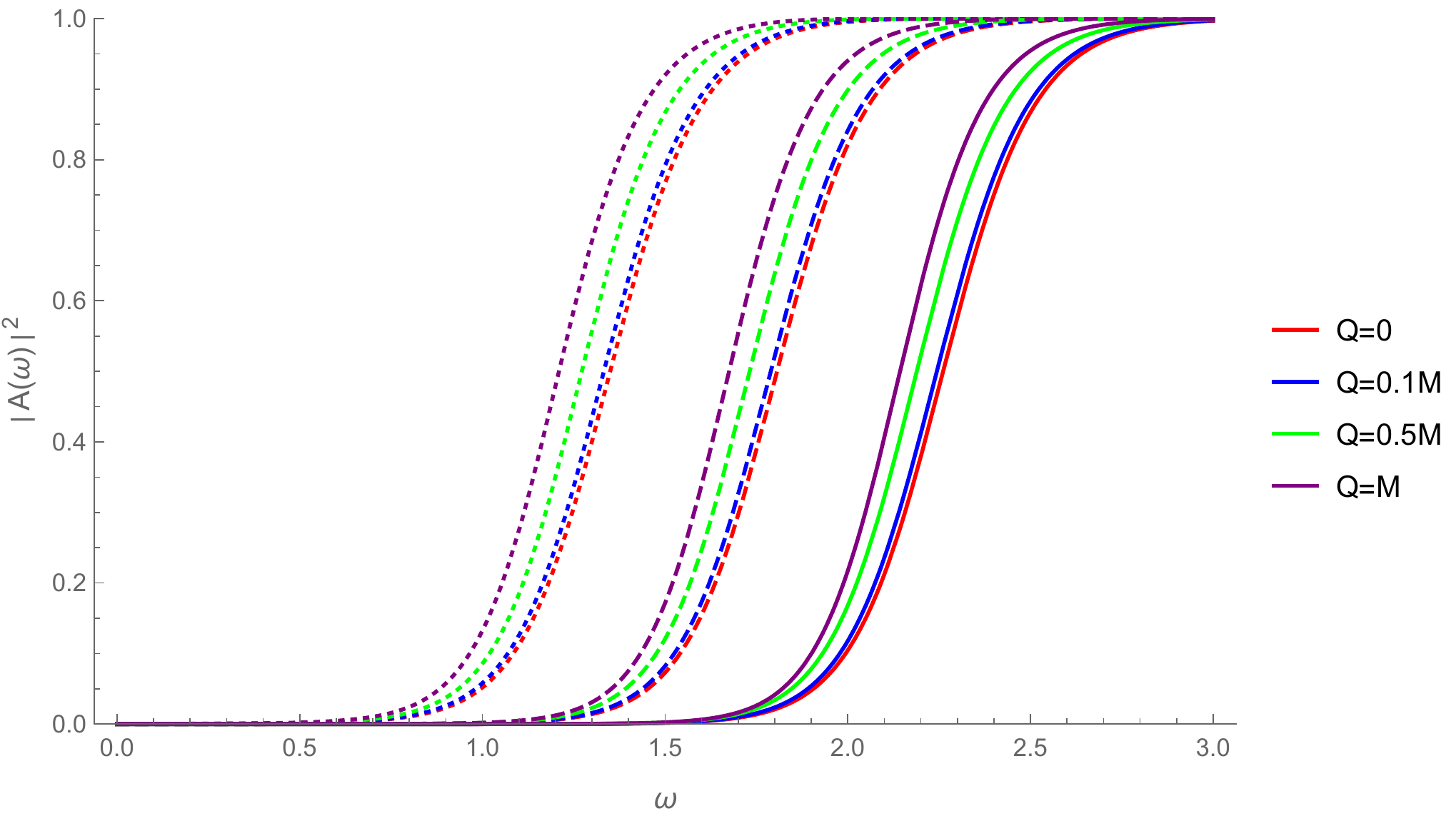}
\caption{$D=6$}
\label{Fig:AP25}
\end{subfigure}
\begin{subfigure}{0.48\textwidth}
\includegraphics[width=\textwidth]{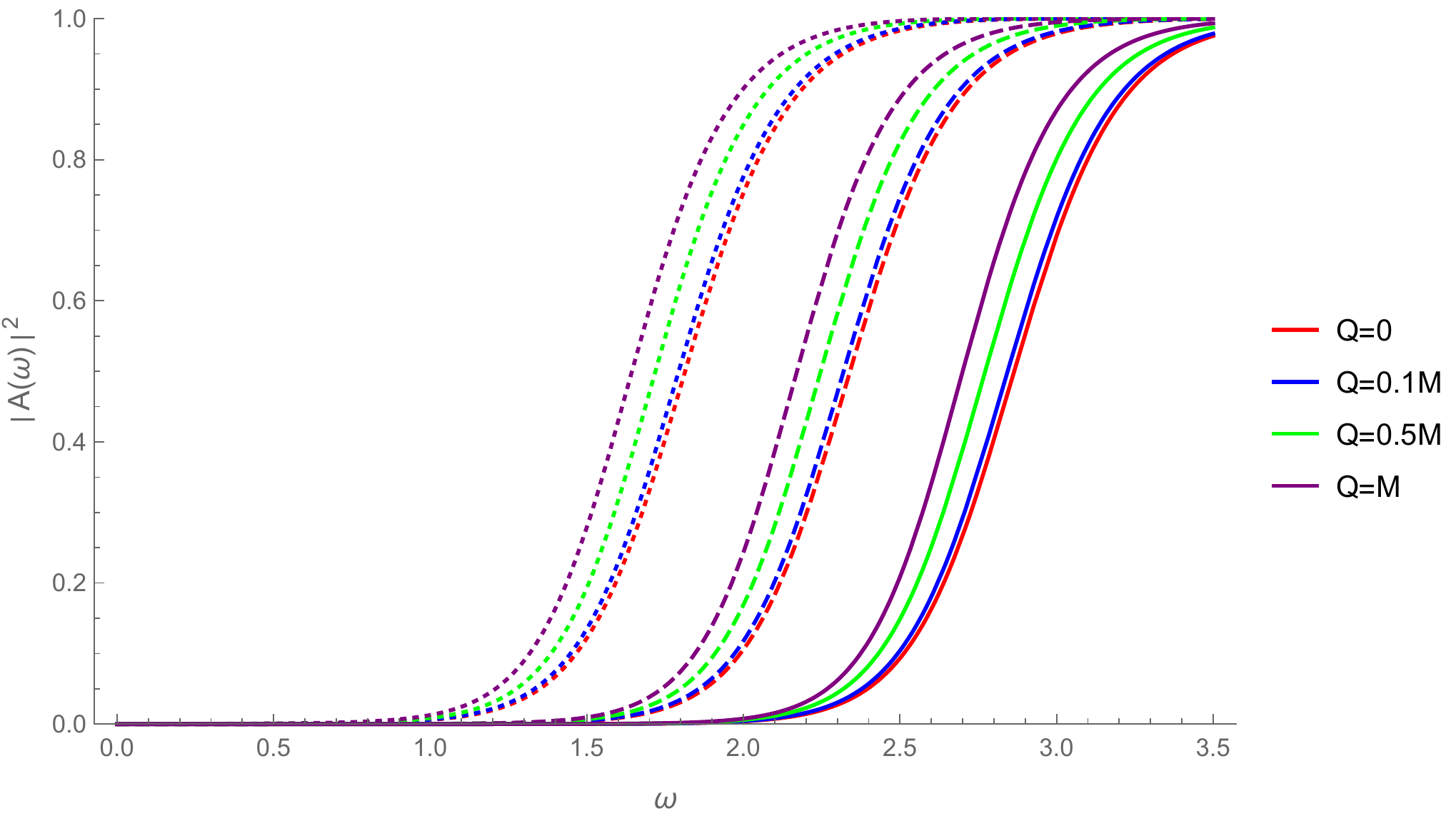}
\caption{$D=7$}
\label{Fig:AP25}
\end{subfigure}
\caption{\it Spin-3/2 field absorption probabilities with various dimensions and $j=3/2$ (Dotted), $j=5/2$ (Dashed) and $j=7/2$ (Solid).}
\label{Fig:APHDtt}
\end{figure}

\FloatBarrier
The absorption probabilities associated with the ``TT eigenmodes" are present in Fig.~\ref{Fig:APHDtt}. It is clear that the absorption probabilities shift from lower energy to higher energy when we increase $j$ (with fixed $Q$ and $D$), and when we increase $D$ (with fixed $Q$ and $j$). However, when $Q$ is increased with fixed $D$ and $j$, the absorption probabilities shift from higher energy to lower energy. This is due to the maximum value of effective potential decreasing when $Q$ increases, as shown in Fig.~\ref{TTPotential} for the typical case of $D=5$, $j=5/2$.

\FloatBarrier
\section{Discussion and conclusion}\label{Sec:Conclusion}

\par In this paper we continue with our previous consideration of spin-3/2 fields in higher dimensional spherically symmetry black hole spacetimes \cite{Chen2015,Chen2016}. The first difference we encounter is the modification of the covariant derivative to the supercovariant derivative for spacetimes with a nonvanishing Ricci tensor, in which Reissner-Nordstr\"om black hole spacetime was the example studied here. This modification is necessary to maintain the supersymmetric gauge symmetry. We have not shown it explicitly in this paper, but the same procedure can also be applied to asymptotic non-flat cases, like black holes in de Sitter and anti-de Sitter spaces.

\par Our main results on the QNM frequencies and the absorption probabilities of the spin-3/2 fields for both non-TT eigenmodes and TT eigenmodes are given in Sec.~\ref{Sec:QNMs} and Sec.~\ref{Sec:Abs}, respectively. First we looked at the non-TT modes, where we found that when the charge of the black hole $Q$ is increased from 0 to $M$ the maximum value of the effective potential increases, while the peak of the potential becomes sharper (Fig.~\ref{nTTPOTENTIAL} is a typical example). The result of this on the quasi-normal frequencies is that both the real part and the magnitude of the imaginary part will increase. For the absorption probability the curve will shift to higher energy when $Q$ is increased. However, this trend will be reversed from the $j=3/2$ and $D=7$ case upwards, such that the maximum value of the potential will instead decrease as $Q$ is increased. For higher dimensions, more and more modes would have this behavior.

\par For the TT-modes, the situation seems to be simpler. Firstly we need to mention that the effective potential in this case is not the same as the one for Dirac fields in the same spacetime, this is due to the extra terms present in the supercovariant derivative. A typical example of the change of the potential with changing $Q$ is given in Fig.~\ref{TTPotential}. We can see that when $Q$ is increased, the maximum value of the potential decreases and the peak broadens. Hence, the corresponding real part of the quasi-normal frequency will decrease, and so too the magnitude of the imaginary part. For the absorption probability the curve will then shift to lower energies as $Q$ is increased. This is opposite to the trend observed for the non-TT cases when the dimension is $D<7$.

\par We have found that for higher dimensions, and especially for the charge $Q$ near the extremal value, the effective potential will develop another maximum. We believe that this is also a property of the potentials in high enough dimensions. The shape of the potential will become more complicated due to the appearance of more maxima and minima. This will pose difficulties to the WKB approximations and the AIM we used to evaluate the QNMs, as well as the absorption probabilities. This problem is more prominent for larger values of $Q$, especially for the extremal cases.

\par Since our method is applicable to spherically symmetric spacetimes, the immediate applications would be to consider spin-3/2 fields for Schwarzschild and Reissner-Nordstr\"om black holes in de Sitter and anti-de Sitter spaces. Charged black holes in anti-de Sitter spaces are particularly interesting because of their relevance to the ground state of supergravity. We are also interested in working out the absorption cross-sections in our subsequent works. To do that we need to find the degeneracies of the eigenspinor-vectors on the $N$-sphere. One should be able to do that by following the method of Camporesi and Higuchi developed for Dirac spinors \cite{camhig}. 


\section*{Acknowledgements}

ASC and GEH are supported in part by the National Research Foundation of South Africa. CHC and HTC are supported in part by the Ministry of Science and Technology, Taiwan, ROC under the Grant No. 105-2112-M-032-004. HTC is also supported in part by the National Center for Theoretical Sciences (NCTS).



\begin{thebibliography}{99}

\bibitem{dasfre}
A.~Das and D.~Z.~Freedman, Nucl. Phys. B {\bf 114}, 271 (1976).

\bibitem{gripen}
M.~T.~Grisaru, H.~N.~Pendelton, and P.~van Nieuwenhuizen, Phys. Rev. D {\bf 15}, 996 (1977).

\bibitem{Chen2015}
C.~H.~Chen, H.~T.~Cho, A.~S.~Cornell, G.~Harmsen, and W.~Naylor, Chin. J. Phys. {\bf 53}, 110101 (2015).

\bibitem{Chen2016}
C.~H.~Chen, H.~T.~Cho, A.~S.~Cornell, and G.~Harmsen, Phys. Rev. D {\bf 94}, 044052 (2016).

\bibitem{konzen}
R.~A. Konoplya and A.~Zhidenko, Rev. Mod. Phys. {\bf 83}, 793 (2011).

\bibitem{kon}
R.~A.~Konoplya, Phys. Rev. D {\bf 68}, 024018 (2003).

\bibitem{chocor1}
H.~T.~Cho, A.~S.~Cornell, Jason Doukas, T.-R.~Huang, and Wade Naylor, Adv. Math. Phys. {\bf 2012}, 281705 (2012).

\bibitem{Liu2014}
J.~T.~Liu, L.~A.~P.~Zayas, and Z.~Yang, J. High Energy Phys. {\bf 1402}, 095 (2014).

\bibitem{KodaIshi2004}
H.~Kodama and A.~Ishibashi, Prog. Theor. Phys. {\bf 111}, 29 (2004)

\bibitem{cho2007split}
H.~T.~Cho, A.~S.~Cornell, Jason Doukas, and Wade Naylor, Phys. Rev. D {\bf 75}, 104005 (2007).

\bibitem{iyewil}
S.~Iyer and C.~M.~Will, Phys. Rev. D {\bf 35}, 3621 (1987).

\bibitem{camhig}
R.~Camporesi and A.~Higuchi, J. Geom. Phys. {\bf 20}, 1 (1996).

\end{thebibliography}
\end{document}